\documentclass[journal]{IEEEtran}
\usepackage[utf8]{inputenc}
\usepackage{biblatex}
\usepackage{amsmath}
\usepackage{amssymb}
\usepackage{systeme}
\usepackage{graphicx}
\usepackage{caption}
\usepackage{cleveref}
\usepackage{bm}
\usepackage[linesnumbered,ruled,vlined]{algorithm2e}
\usepackage{amsthm}
\usepackage{spalign}
\usepackage{amsfonts}
\usepackage{enumitem}

\graphicspath{ {./images/} }

\let\oldnl\nl
\newcommand{\nonl}{\renewcommand{\nl}{\let\nl\oldnl}}

\newtheorem{theorem}{Theorem}
\newtheorem{lemma}{Lemma}

\newcommand{\xAB}{\mathbf{x}_{A \rightarrow B}}
\newcommand{\xBA}{\mathbf{x}_{B \rightarrow A}}

\newcommand{\vAB}{v_{A \rightarrow B}}
\newcommand{\vBA}{v_{B \rightarrow A}}

\newcommand{\qAB}{q_{A \rightarrow B}}
\newcommand{\qBA}{q_{B \rightarrow A}}

\newcommand{\hvAB}{\Hat{v}_{A \rightarrow B}}
\newcommand{\hvBA}{\Hat{v}_{B \rightarrow A}}

\newcommand{\hvw}{\Hat{v}_w}
\newcommand{\hgam}{\Hat{\gamma}}

\newcommand{\heta}{\Hat{\eta}}
\newcommand{\hzeta}{\Hat{\zeta}}

\newcommand{\MSE}{\mathcal{E}}

\newcommand{\by}{\mathbf{y}}

\newcommand{\bx}{\mathbf{x}}
\newcommand{\bw}{\mathbf{w}}
\newcommand{\bh}{\mathbf{h}}
\newcommand{\bq}{\mathbf{q}}
\newcommand{\be}{\mathbf{e}}
\newcommand{\br}{\mathbf{r}}
\newcommand{\bp}{\mathbf{p}}
\newcommand{\bz}{\mathbf{z}}

\newcommand{\bQ}{\mathbf{Q}}
\newcommand{\bv}{\mathbf{v}}
\newcommand{\bd}{\mathbf{d}}
\newcommand{\bff}{\mathbf{f}}
\newcommand{\bg}{\mathbf{g}}

\newcommand{\bzero}{\mathbf{0}}
\newcommand{\bmu}{\bm{\mu}}
\newcommand{\bLambda}{\bm{\Lambda}}
\newcommand{\bA}{\mathbf{A}}
\newcommand{\bI}{\mathbf{I}}
\newcommand{\bU}{\mathbf{U}}
\newcommand{\bV}{\mathbf{V}}
\newcommand{\bS}{\mathbf{S}}
\newcommand{\bW}{\mathbf{W}}
\newcommand{\bH}{\mathbf{H}}

\newcommand{\bF}{\mathbf{F}}
\newcommand{\bZ}{\mathbf{Z}}

\newcommand{\bC}{\mathbf{C}}

\newcommand{\bP}{\mathbf{P}}

\newcommand{\bD}{\mathbf{D}}
\newcommand{\bM}{\mathbf{M}}

\newcommand{\bG}{\mathbf{G}}
\newcommand{\bmm}{\mathbf{m}}
\newcommand{\bgamma}{\bm{\gamma}}
\newcommand{\bPsi}{\bm{\Psi}}
\newcommand{\bPhi}{\bm{\Phi}}

\newcommand{\bphi}{\bm{\phi}}
\newcommand{\bsigma}{\bm{\sigma}}
\newcommand{\bSigma}{\bm{\Sigma}}

\newcommand{\bone}{\bm{1}}

\newcommand{\bhW}{\hat{\bW}}

\newcommand{\normDensity}{\mathcal{N}}

\newcommand{\limN}{\lim_{N \rightarrow \infty}}
\newcommand{\as}{\overset{a.s.}{=}}
\newcommand{\inv}[1]{\frac{1}{#1}}

\SetKwInput{KwInput}{Input}               
\SetKwInput{KwOutput}{Output}
\SetKwInOut{KwInitialization}{Initialization}
\SetKwBlock{KwBlockA}{Block A}{end}
\SetKwBlock{KwBlockB}{Block B}{end}

\SetKwBlock{KwBlockCGRoutine}{Standard CG routine}{end}
\SetKwBlock{KwBlockCGDivergence}{CG correction}{end}
\SetKwBlock{KwBlockCGMSE}{SE of Block A}{end}

\begin{document}

\title{Compressed Sensing with Upscaled Vector Approximate Message Passing}

\author{Nikolajs Skuratovs, 
Michael Davies, ~\IEEEmembership{Fellow IEEE}
\thanks{This work was supported by the ERC project C-SENSE (ERC-ADG-2015-694888). MD is also supported by a Royal Society Wolfson Research Merit Award.

Copyright (c) 2017 IEEE. Personal use of this material is permitted.  However, permission to use this material for any other purposes must be obtained from the IEEE by sending a request to pubs-permissions@ieee.org.
}}

\maketitle

\begin{abstract}
    The Recently proposed Vector Approximate Message Passing (VAMP) algorithm demonstrates a great reconstruction potential at solving compressed sensing related linear inverse problems. VAMP provides high per-iteration improvement, can utilize powerful denoisers like BM3D, has rigorously defined dynamics and is able to recover signals measured by highly undersampled and ill-conditioned linear operators. Yet, its applicability is limited to relatively small problem sizes due to the necessity to compute the expensive LMMSE estimator at each iteration. In this work we consider the problem of upscaling VAMP by utilizing Conjugate Gradient (CG) to approximate the intractable LMMSE estimator. We propose a rigorous method for correcting and tuning CG withing CG-VAMP to achieve a stable and efficient reconstruction. To further improve the performance of CG-VAMP, we design a warm-starting scheme for CG and develop theoretical models for the Onsager correction and the State Evolution of Warm-Started CG-VAMP (WS-CG-VAMP). Additionally, we develop robust and accurate methods for implementing the WS-CG-VAMP algorithm. The numerical experiments on large-scale image reconstruction problems demonstrate that WS-CG-VAMP requires much fewer CG iterations compared to CG-VAMP to achieve the same or superior level of reconstruction.
\end{abstract}


\section{Introduction}

In this work we focus on reconstructing a random signal $\bx \in \mathbb{R}^N$ from the following system of measurements 

\noindent
\begin{equation}
    \by = \bA \bx + \bw
    \label{eq:y_measurements}
\end{equation}

\noindent
where $\by \in \mathbb{R}^M$, $\bw \in \mathbb{R}^M$  is a zero-mean i.i.d. Gaussian noise vector $\bw \sim N(0,v_w \bI_M)$ and $\bA \in \mathbb{R}^{M \times N}$ is a measurement matrix that is assumed to be available. We consider the large-scale compressed sensing scenario where $M < N$ with the ratio $\frac{M}{N} = \delta = O(1)$. Additionally, we are interested in the cases where the operator $\bA$ might be ill-conditioned. This particular setting is common for many computational imaging applications including CT and MRI \cite{CT_paper_1}, \cite{MRI_paper_1}, \cite{Shniter_MRI}.

The problem of recovering $\bx$ from $\by$ as in \eqref{eq:y_measurements} can be approached by many first-order methods including Iterative Soft-Thresholding (IST) \cite{ISTA} and Approximate Message Passing (AMP) \cite{AMP}. While algorithmically the original form of AMP and IST are similar, AMP includes an additional term called the \textit{Onsager correction}, which leads to fast convergence rates, low per-iteration cost and high reconstruction quality given the measurement operator $\bA$ is sub-Gaussian \cite{AMP_graphical_models}. Additionally, AMP was shown \cite{D-AMP} to perform very well for natural image reconstruction by utilizing State-of-The-Art denoisers like Non-Local Means \cite{NLM}, BM3D \cite{BM3D}, denoising CNN \cite{D-AMP} etc. Moreover, the asymptotic dynamics of AMP can be fully characterized through a 1D \textit{State Evolution (SE)} \cite{SE_AMP} that establishes the evolution of the Mean Squared Error (MSE) $v_t$ of the iteratively updated estimate $\hat{x}_t$

\noindent
\begin{equation}
    \lim_{N \rightarrow \infty} v_{t+1} \overset{a.s.}{=} SE_t \big( v_{t} \big)
    \label{eq:AMP_SE}
\end{equation}

\noindent
where $t$ stands for the iteration number. This result makes the analysis of stability and efficiency of the resulting algorithm tractable and provides the means of optimizing each step of the algorithm with respect to the MSE. In particular, the SE was used to prove the Bayes-optimal reconstruction results of AMP given the denoiser used in the algorithm is Bayes-optimal \cite{SE_AMP}, \cite{AMP_Bayes_ioptimality_1}, \cite{AMP_Bayes_ioptimality_2}.

Despite the advantages of AMP, it is evidenced that when $\bA$ is not from the family of i.i.d. sub-Gaussian matrices, is even mildly ill-conditioned or is not centered, AMP becomes unstable \cite{VAMP} \cite{AMP_convergence_general_A}, \cite{UnifiedSE}. To derive an algorithm that is more robust with respect to the structure of $\bA$, while preserving the advantages of AMP, we refer to the Bayesian framework based on \textit{Expectation Propagation (EP)} \cite{Minka_EP} that aims to find an approximation to the MMSE estimate of $\bx$ in the iterative manner. In the context of linear inverse problems, EP has been used in many works including \cite{EP_unmixing_Altmann}, \cite{Ko_EP}, \cite{seeger_EP}, \cite{OAMP} and others, but the first work that rigorously established the error dynamics of EP was \cite{VAMP} and shortly after the work \cite{EP_Keigo}. In \cite{VAMP}, EP was used to derive a more general framework called \textit{Vector Approximate Message Passing (VAMP)} whose asymptotic dynamics can be described through a 1D SE given the measurement operator is right-orthogonally invariant, i.e. in the SVD\footnote{In this work we will broadly use the Singular Value Decomposition (SVD) of the measurement matrix $\bA = \bU \bS \bV^T$ as a tool to prove certain key statements and discuss important properties of algorithms. However, we will never use SVD in the algorithms.} of $\bA = \bU \bS \bV^T$ the right singular-vector matrix $\bV$ is Haar distributed \cite{rand_mat_methods_book}. VAMP demonstrates high per-iteration improvement by utilizing powerful denoisers like BM3D \cite{BM3D} and was shown to achieve Bayes-optimal reconstruction results given the denoiser is Bayes-optimal and conditioned on the fact that the replica prediction for ROI matrices is exact \cite{VAMP}. Importantly, these results hold for any singular spectrum of $\bA$ as long as the emperical eigenvalue distribution \cite{rand_mat_methods_book} of $\bA \bA^T$ converges to a random variable with a compact support \cite{VAMP}, \cite{EP_Keigo}. 

Yet, applicability of VAMP is limited to relatively small dimensional problems due to the necessity to compute the solution to a system of linear equations (SLE) of dimension $M$, which can be intractable for large dimensional data as in the modern imaging problems. While it was shown \cite{VAMP} that precomputing the SVD can help with this, in such applications storing the singular vector matrices may also be intractable from the memory point of view. The computational/memory bottleneck of VAMP was partially addressed in \cite{CG_EP}, where it was suggested to approximate the exact solution by the Conjugate Gradient (CG) algorithm and correct the obtained approximation to preserve stability of the algorithm and the 1D SE. However, the proposed methods of correcting the CG algorithm and estimating the SE require the access to the moments of the singular spectrum $\bS \bS^T$ up to the order $2i+2$, where $i$ is the maximum number CG iterations. In practice the singular spectrum is not available, while using the estimated moments leads to unstable performance of the algorithm even with a small $i$.

In this work we propose scalable tools for correcting, tuning and accelerating the CG algorithm within CG-VAMP to approach its performance towards VAMP's while being able to address large scale linear inverse problems in reasonable and controllable time. First, we develop a rigorous asymptotic model of the correction for the CG algorithm within CG-VAMP and use the model to construct a robust and prior-free method for estimating this correction. The proposed method can be easily implemented in the body of CG to correct it after each inner-loop iteration, it has a negligible additional computational and memory costs, it does not rely on any prior information about the singular spectrum of $\bA$ and remains stable for a much larger range of $i$ compared to the method proposed in \cite{CG_EP}

Second, we develop an Adaptive CG (ACG) algorithm that selects the number of the inner-loop iterations $i$ that guarantees the desired improvement level of the SE after each outer-loop iteration $t$. The developed stopping criteria leverage the proposed rigorous SE model of CG and ensure consistent progression and balanced operation of the whole CG-VAMP algorithm. The simulations provided demonstrate high accuracy of reconstruction, efficiency and stability of the algorithm.

The third contribution is the proposed CG warm-starting scheme that leads to the improved dynamics of the resulting Warm-Started CG-VAMP (WS-CG-VAMP). We provide the intuitive motivation for warm-starting, rigorously adapt the structure of CG-VAMP to use Warm-Started CG (WS-CG) and develop the rigorous multi-dimensional SE of the WS-CG-VAMP algorithm. Additionally, we provide an asymptotic model of the correction for WS-CG and propose a prior-free and robust method for estimating it within WS-CG-VAMP. The numerical experiments demonstrate that when both CG-VAMP and WS-CG-VAMP algorithms are provided with the same amount of resources, the later algorithm converges much closer to the VAMP fixed point.

\subsection{Notations}

We use roman $v$ for scalars, small boldface $\bv$ for vectors and capital boldface $\bV$ for matrices. We use the identity matrix $\bI_N$ with a subscript to define that this identity matrix is of dimension $N$ or without a subscript where the dimensionality is clear from the context. We define $\bzero_k$ and $\bone_k$ to be column vectors of dimension $k$ of zeros and of ones respectively and use $(\bx)_n$ to define $n$-th element of a vector $\bx$. By $||\bx||_p$ we mean the $l_p$ norm of the vector $\bx$ and we usually omit the subscript $2$ for the $l_2$ norm so $||\bx||_2$ and $||\bx||$ refer to the same operation. We define $Tr\big\{ \bM \big\}$ to be the trace of a matrix $\bM$ and $\kappa(\bM)$ to be the condition number of $\bM$. The divergence of a function $f(\bx)$ with respect to the vector $\bx$ is defined as $\nabla_{\bx} \cdot f(\bx) = \sum_{i = 1}^N \frac{f(\bx)}{\partial x_i}$. By writing $q(\bx) = \normDensity(\bx; \bmm, \bSigma)$ we mean that the density $q(\bx)$ is normal with the mean vector $\bmm$ and the covariance matrix $\bSigma$. Additionally, we use $\bx \sim q(\bx)$ to indicate that the random vector $\bx$ is distributed according to a density $q$ and $\by \sim \bx$ to indicate that the vector (or a scalar) $\by$ has the same distribution as $\bx$. We reserve the letter $t$ for the outer-loop iteration number of the EP- and VAMP-based algorithms and use the letter $i$ for the iteration number of the Conjugate Gradient algorithm. Lastly, we mark an estimate of some values such as $v$ with a hat, i.e. $\Hat{v}$.

\section{Expectation Propagation framework and its extensions}

In this section we review the main framework -- Expectation Propagation (EP) \cite{Minka_EP} -- that will be used as a basis for our work. Additionally, we discuss what methods can be used to generalize EP and analyze their properties and limitations.

\subsection{Background on Expectation Propagation} \label{sec:EP_background}

The main goal of EP is the construction of a Gaussian factorized density approximating the original factorized density by means of matching the two in terms of Kullback-Leibler (KL) divergence. In our case, we aim to approximate the posterior density

\noindent
\begin{equation}
    p(\bx|\by) \propto p(\by|\bx) p(\bx)
    \label{eq:posterior}
\end{equation}

\noindent
as a product of two isotropic Gaussian\footnote{While here we aim to approximate the likelihood and the prior with isotropic Gaussian densities, the original EP framework allows to construct approximations from the exponential family, where the isotropic Gaussian family is a special case.} densities

\noindent
\begin{gather}
    \qAB(\bx) = \normDensity\big( \xAB, \vAB \bI \big) \label{eq:EP_qAB_Gaussian}\\
    \qBA(\bx) = \normDensity\big( \xBA, \vBA \bI \big) \label{eq:EP_qBA_Gaussian}
\end{gather}

\noindent
that approximate the likelihood and the prior densities respectively. For the simple factorization as in \eqref{eq:posterior}, the EP update rules \cite{Minka_EP} result in the following iterative process

\noindent
\begin{gather}
    \qAB(\bx) = \frac{proj\big[ \qBA(\bx) p(\by|\bx) \big]}{\qBA(\bx)} = \frac{proj\big[ q_A(\bx) \big]}{\qBA(\bx)} \label{eq:EP_qAB}\\
    \qBA(\bx) = \frac{proj\big[ \qAB(\bx) p(\bx) \big]}{\qAB(\bx)} = \frac{proj\big[ q_B(\bx) \big]}{\qBA(\bx)} \label{eq:EP_qBA}
\end{gather}

\noindent
where $proj[ \cdot ]$ is the KL projection operator on the family of isotropic Gaussian densities \cite{EP_for_lin_model}. Although in general this projection operator can produce a Gaussian density with arbitrary covariance matrix, computing the full covariance matrix might be either impossible or slow \cite{EP_unmixing_Altmann}. Therefore we stick to the isotropic Gaussian family.

The iterative refinement process \eqref{eq:EP_qAB}-\eqref{eq:EP_qBA} is split into two symmetric blocks: Block A and Block B that are associated with the likelihood density $p(\by|\bx)$ and with the prior density $p(\bx)$ respectively. For example, in Block A we update the so-called \textit{tilted density} $q_A$, which acts as a new approximation to the posterior density \eqref{eq:posterior}, based on the \textit{cavity density} $\qBA(\bx)$ that represents the information sent from Block B to Block A. Next, we compute the new cavity density $\qAB(\bx)$ that acts as a likelihood approximation and is sent to Block B where the same process repeats. 

Due to the nature of the cavity densities \eqref{eq:EP_qAB_Gaussian} and \eqref{eq:EP_qBA_Gaussian}, the only information required from the tilted densities $q_A$ and $q_B$

\noindent
\begin{gather}
    q_A(\bx) = \qBA(\bx) p(\by|\bx) \label{eq:tilde_A}\\
    q_B(\bx) = \qAB(\bx) p(\bx) \label{eq:tilde_B}
\end{gather}

\noindent
are the mean vectors $\bmu_A$ and $\bmu_B$ and the variances $v_A$ and $v_B$. Then, with these notations, the update rules \eqref{eq:EP_qAB}-\eqref{eq:EP_qBA} are equivalent to

\noindent
\begin{gather}
    \qAB(\bx) = \frac{N \big( \bx; \bmu_A, v_A \big)}{N\big( \bx; \xBA, \vBA \big)} \label{eq:EP_frac_AB}\\
    \qBA(\bx) = \frac{N \big( \bx; \bmu_B, v_B \big)}{N\big( \bx; \xAB, \vAB \big)} \label{eq:EP_frac_BA}
\end{gather}

\noindent
If one derives the update rules for the mean vector $\xAB$ and $\xBA$ and the variances $\vAB$ and $\vBA$ based on \eqref{eq:tilde_A}-\eqref{eq:EP_frac_BA} and after some rearrangements suggested in \textit{Appendix B} in \cite{EP_Keigo}, one can obtain the algorithm shown in Algorithm 1. Since in practice we usually do not have the access to the true values of the variances $v_x \as \limN \inv{N} ||\bx||^2$ and $v_w \as \limN \inv{M} ||\bw||^2$ of the true signal $\bx$ and of the noise measurement $\bw$ respectively, the algorithm uses their estimated versions $\Hat{v}_x$ and $\Hat{v}_w$. 

This algorithm is naturally split into two parts: Block A and Block B, where we compute the approximations for the likelihood and the prior respectively, while the arrows mean the information send from one block to another. Each block contains a function $\bg_{(\cdot)}$ that computes a vector associated with the mean\footnote{While in Block B the function $\bg_B$ aims to compute the mean of the density $q_B$, the function $\bg_A$ computes a vector that is related to the mean $\bmu_A$ through Woodbury transformation. Originally \cite{VAMP}, the mean $\bmu_A$ corresponds to the LMMSE estimator computed in the signal's domain of dimension $N$, while in Algorithm 1 we compute a similar 'LMMSE' estimator in the measurement domain of dimension $M$, which is significantly cheaper. For more details, please refer to \textit{Appendix B} in \cite{EP_Keigo}.} of the densities $q_A$ and $q_B$. As it will be shown further, computation of the variances $v_A$ and $v_B$ of the densities $q_A$ and $q_B$ is not required for the algorithm to iterate and, therefore, their computation is omitted in Algorithm 1. The mean estimates $\xAB$ and $\xBA$ and the estimates of the variances $\hvAB$ and $\hvBA$ of the two approximating densities $\qAB$ and $\qBA$ are derived based on the ratios from \eqref{eq:EP_frac_AB} and \eqref{eq:EP_frac_BA}.

\begin{algorithm}
\DontPrintSemicolon
\SetNoFillComment
\caption{EP-based Algorithm}

    \KwInitialization{ $\xBA^0 = 0, \hvBA^0 = \Hat{v}_x, t=0$}

    \While{$t < T_{max}$ and $\hvBA^t \geq \epsilon$}
    {
        \KwBlockA
        {
            $\bmu_A^t = \bg_A(\xBA^t, \hvBA^t)$ \;
   	        $\hgam_A^t = \frac{1}{N} \nabla_{(\xBA^t)} \cdot \Big( \bA^T \bg_A(\xBA^t, \hvBA^t) \Big)$ \;
   	        $\xAB^t = \xBA^t - \inv{\hgam_A^t} \bA^T \bmu_A^t$ \;
   	        $\hvAB^t = f_A(\hvBA^t)$
        }
        
        \KwBlockB
        {
            $\bmu_B^{t+1} = \bg_B(\xAB^t,\hvAB^t)$ \;
            $\hgam_B^{t+1} = \frac{1}{N} \nabla_{(\xAB^t)} \cdot \bg_B(\xAB^t,\hvAB^t) $ \;
            $\xBA^{t+1} = \frac{1}{1-\ \hgam_B^{t+1}} \Big( \bmu_B^{t+1} - \hgam_B^{t+1} \xAB^t \Big) $ \;
            $\hvBA^{t+1} = f_B(\hvAB^t)$ \;
        }
        $t = t + 1$ \;
    }
    \KwOutput{$\bmu_B^t$}    
\end{algorithm}

\subsection{The Large System Limit model of EP}

Although the original EP update rules that are a special case of Algorithm 1 were derived before the works \cite{OAMP}, \cite{VAMP} and \cite{EP_Keigo}, in the latter two works the authors rigorously studied the dynamics of the EP algorithm and analyzed how the error propagates throughout iterations. Their analysis is based on three main assumptions: 

\textbf{Assumption 1}: The measurement matrix $\bA$ is Right-Orthogonally invariant (ROI) such that in the SVD of $\bA = \bU \bS \bV^T$, the matrix $\bV$ is independent of other random terms and is Haar distributed \cite{rand_mat_methods_book}.

\textbf{Assumption 2}: The dimensions of the signal model $N$ and $M$ approach infinity with a fixed ratio $\delta = \frac{M}{N}$ and in the Large System Limit the spectrum of the eigenvalues of the matrix $\bS \bS^T$ almost surely converges to a distribution with a compact support.

\textbf{Assumption 3}: The function $\bg_B$ is uniformly Lipschitz so that the sequence of functions $\bg_B: \mathbb{R}^N \mapsto \mathbb{R}^N$ indexed by $N$ are Lipschitz continuous with a Lipschitz constant $L_N < \infty$ as $N \rightarrow \infty$ \cite{NS-VAMP}, \cite{AMP_SE_non_separable}. Additionally, we assume the sequences of the following inner-products exist and are almost surely finite as $N \rightarrow \infty$ \cite{NS-VAMP}

\noindent
\begin{gather*}
    \lim_{N \rightarrow \infty} \inv{N} \bg_B(\bx + \bd_1)^T \bg_B(\bx + \bd_2), \; \; \lim_{N \rightarrow \infty} \inv{N} \bx^T \bg_B(\bx + \bd_1), \\ 
    \lim_{N \rightarrow \infty} \inv{N} \bd_1^T \bg_B(\bx + \bd_2), \quad \lim_{N \rightarrow \infty} \inv{N} \bx^T \bd_1, \quad  \lim_{N \rightarrow \infty} \inv{N} ||\bx||^2 
\end{gather*}

\noindent
where $\bd_1, \bd_2 \in \mathbb{R}^N$ with $( (\bd_1)_n, (\bd_2)_n ) \sim \normDensity(\bzero_2, \bC)$ for some positive definite $\bC \in \mathbb{R}^2$. 

Additionally, without loss of generality, we assume the normalization $\inv{N} Tr\{\bA \bA^T\} = 1$. Then, under such assumptions, the mean vectors $\xAB^t$ and $\xBA^t$ of the cavity densities $\qAB^t$ and $\qBA^t$ can be modeled as \cite{VAMP}, \cite{EP_Keigo} 

\noindent
\begin{gather}
    \xBA^t = \bx + \bq_t \label{eq:x_ba} \\
    \xAB^t = \bx + \bh_t \label{eq:x_ab}
\end{gather}

\noindent
where the error vectors $\bq_t$ and $\bh_t$ are zero-mean with the variance $\vBA^t$ and $\vAB^t$ respectively. The key result proved in both \cite{VAMP} and \cite{EP_Keigo} is that the error vector $\bh_t \in \mathbb{R}^N$ is asymptotically zero-mean i.i.d Gaussian and is independent of $\bq_{\tau}$ for $\tau \leq t$ 

\noindent
\begin{gather}
    \lim_{N \rightarrow \infty} (\bh_t)_n \overset{a.s.}{\sim} \normDensity(0, \vAB^t) \label{eq:h_t_gaussian} \\
    \lim_{N \rightarrow \infty} \inv{N} \bh_t^T \bq_{\tau} \overset{a.s.}{=} 0 \label{eq:h_q_independence} 
 \end{gather}

\noindent
Note that Algorithm 1 is initialized with $\xBA^t = \bzero_N$ so we have $\bq_0 = -\bx$, which together with \eqref{eq:h_q_independence} implies that $\bh_t$ is asymptotically independent of $\bx$. Additionally, the result \eqref{eq:h_t_gaussian} together with the definition of $q_B$ indicate that, in the limit, the mean estimator $\bg_B$ operating on the intrinsic measurements $\xAB^t$ performs Minimum MSE (MMSE) estimation of $\bx$ from a Gaussian channel. For that reason, the function $\bg_B$ is further referred as the \textit{denoiser}. Similarly, one can use the definition of the likelihood $p(\by|\bx)$ and of the cavity density $\qBA$ to show that in EP the function $\bg_A$ corresponds to the following linear mapping \cite{EP_Keigo}

\noindent
\begin{equation}
    \bg_A(\xBA^t, \hvBA^t) = \bW_t^{-1} \big( \by - \bA \xBA^t \big) \label{eq:LMMSE}
\end{equation}

\noindent
where

\noindent
\begin{equation}
        \bW_t = v_w \bI_M + \vBA^t \bA \bA^T \label{eq:W_t}
\end{equation}

\noindent
Due to the structure of the density $q_A$, the estimator $\bg_A$ from \eqref{eq:LMMSE} is referred to as the \textit{Linear Minimum Mean Squared Error (LMMSE)} estimator. In the following we will use \eqref{eq:W_t} for the analysis of EP dynamics, while in practice we implement $\bhW_t = \hvw \bI_M + \hvBA^t \bA \bA^T$, where $\hvw$ and $\hvBA^t$ are the estimated values of the true variances $v_w$ and $\vAB^t$.

The key ingredients ensuring the main asymptotic properties of EP are the scalars $\gamma_A^t$ and $\gamma_B^t$ that are explicitly designed to impose the orthogonality \eqref{eq:h_q_independence}. By substituting the update for $\xAB^t$ from Algorithm 1 into \eqref{eq:x_ab}, one can confirm that, for $\bq_t = \xBA^t - \bx$ and $\bh_t = \xAB^t - \bx$ to be orthogonal after the update in Block A, the scalar $\gamma_A^t$ must follow \cite{UnifiedSE}

\noindent
\begin{equation}
    \limN \inv{N} \bq_t^T \bA^T \bg_A(\xBA^t, \hvBA^t) \as \gamma_A^t \vBA^t \label{eq:gamma_A_general_equality}
\end{equation}

\noindent
Similarly, the scalar $\gamma_B^t$ ensuring \eqref{eq:h_q_independence} after the update in Block B must follow \cite{UnifiedSE}

\noindent
\begin{equation}
    \limN \inv{N} \bh_t^T \bg_B(\xAB^t,\hvAB^t) \as \gamma_B^t \vAB^t \label{eq:gamma_B_general_equality}
\end{equation}

\noindent
Unfortunately, these identities can only be used for theoretical purposes and not be implemented in practice since they are explicitly formulated in terms of the error vectors $\bq_t$ and $\bh_t$. The important result shown in \cite{VAMP}, \cite{EP_Keigo} and other works on Message Passing is that in the LSL, \eqref{eq:gamma_A_general_equality} and \eqref{eq:gamma_B_general_equality} converge to the asymptotic normalized divergences of estimators $\bg_A$ and $\bg_B$ respectively \cite{VAMP}

\noindent
\begin{gather}
    \gamma_A^t \as \lim_{N \rightarrow \infty} \inv{N} \nabla_{\xBA^t} \cdot \Big( \bA^T \bg_A(\xBA^t, \hvBA^t) \Big) \label{eq:gamma_A_general} \\
    \gamma_B^{t+1} \as \lim_{N \rightarrow \infty} \inv{N} \nabla_{\xAB^t} \cdot \Big( \bg_B(\xAB^t,\hvAB^t) \Big) \label{eq:gamma_B_general}
\end{gather}

\noindent
which corresponds to their updates in Algorithm 1. Furthermore, for the MMSE estimator $\bg_B$ and the LMMSE estimator $\bg_A$, these scalars have the following closed-form solutions \cite{EP_Keigo}

\noindent
\begin{gather}
    \gamma_B^t \overset{a.s.}{=} \frac{\MSE(\vAB^t)}{\vAB^t} \label{eq:gamma_B_EP}\\
    \gamma_A^t \overset{a.s.}{=} \delta \int \frac{\lambda}{v_w + \vBA^t \lambda} p(\lambda) d\lambda \label{eq:gamma_A_EP}
\end{gather}

\noindent
where $p(\lambda)$ is the limiting spectral distribution of $\bLambda = \bS \bS^T$ and $\MSE(\vAB^t)$ 

\noindent
\begin{gather}
    \MSE(\vAB^t) = \limN \inv{N} ||\bg_B(\bx + \be, \vAB^t) - \bx||^2 \nonumber\\
    \be \sim \normDensity(\bzero_N, \vAB^t \bI_N) \nonumber
\end{gather}

\noindent
is the MMSE conditioned on the intrinsic measurement $\xAB^t$ from \eqref{eq:x_ab} \cite{VAMP}.

Another important asymptotic property of EP under Assumptions 1-3 is that its performance can be fully characterized by a 1D State Evolution (SE) as for AMP. This result is summarized in the following lemma.

\begin{lemma} \label{lemma:VAMPs_SE}

\textit{\cite{VAMP}, \cite{EP_Keigo}:} Let $\bg_A$ be the LMMSE estimator \eqref{eq:LMMSE} and $\bg_B$ be the MMSE denoiser. Then, under Assumptions 1-3, the variances $\vBA^t \overset{a.s.}{=} \lim_{N \rightarrow \infty} \inv{N} || \xBA^t - \bx ||^2$ and $\vAB^t \overset{a.s.}{=} \lim_{N \rightarrow \infty} \inv{N} || \xAB^t - \bx ||^2$ of the intrinsic error vectors $\bq_t = \xBA^t - \bx$ and $\bh_t = \xAB^t - \bx$ evolve as

\noindent
\begin{gather}
    \vAB^t \overset{a.s.}{=} (\gamma_A^t)^{-1} - \vBA^t \label{eq:true_v_ab} \\
    \vBA^{t+1} \overset{a.s.}{=} \frac{\MSE(\vAB^t) - (\gamma_B^t)^2 \vAB^t}{\big( 1 - \gamma_B^t \big)^2} \label{eq:true_v_ba}
\end{gather}

\end{lemma}

\subsection{Approximate EP -- Vector Approximate Message Passing} \label{sec:VAMP}

One of the limitations of the original EP algorithm is that in Block B, the function $\bg_B$ has to compute the MMSE estimate of the signal $\bx$ observed from the Gaussian intrinsic channel $\xAB^t$. Since for most practical problems constructing such an estimator is either impossible or computationally intractable, in \cite{VAMP} the EP algorithm was generalized to operate with any separable Lipschitz function $\bg_B$ and the resulting algorithm was called \textit{Vector Approximate Message Passing (VAMP)}. In the further work \cite{NS-VAMP} the same authors extended the allowed class of denoisers for VAMP to uniformly Lipschitz functions and demonstrated numerically the stability of the algorithm employing denoisers such as BM3D and denoising CNN \cite{D-VAMP}. Equipped with such powerful denoisers, VAMP demonstrates high per-iteration improvement, fast and stable convergence and preserves the SE from Lemma \ref{lemma:VAMPs_SE} \cite{NS-VAMP}. Moreover, in \cite{UnifiedSE} it was shown that a framework like VAMP is flexible with respect to the choice of the estimator $\bg_A$ as well, as long as $\bg_A$ is asymptotically separable. In those cases, the interpretation of the vectors $\bmu_A$ and $\bmu_B$ loses the Bayesian meaning, but instead one obtains a general framework that is proved to be stable and follow the SE as long as the key ingredients -- the divergences of those estimators -- are computed and used in Algorithm 1 \cite{UnifiedSE}.

\subsection{Conjugate Gradient for upscaling VAMP} \label{sec:CG_introduction}

While VAMP exhibits many important advantages, its practical implementation for recovering large-scale data is limited due to intractability of the matrix inverse $\bW_t^{-1}$ in the the LMMSE estimator \eqref{eq:LMMSE}. Although precomputing the SVD of $\bA$ would reduce the computational costs of implementing $\bW_t^{-1}$ \cite{VAMP}, this approach requires storing a matrix of dimension $M$ by $M$, which might be infeasible from the memory point of view. In this work we approach the problem of upscaling VAMP by referring to an iterative method approximating the exact LMMSE solution. In the Message Passing context, this idea was originally considered in the works \cite{CG_EP} and \cite{D-VAMP_poster}, where it was proposed to reformulate the LMMSE estimator as the solution to the system of linear equations (SLE) 

\noindent
\begin{equation}
    \bW_t \bmu_A^t = \bz_t \label{eq:SLE}
\end{equation}

\noindent
and then apply an iterative method for approximating $\bmu_A^t$. While there are various iterative methods suitable for this task, in this work we use the \textit{Conjugate Gradient (CG)} algorithm considered in \cite{CG_EP} and shown in Algorithm 2 as the baseline for upscaling VAMP. On a general level, the CG algorithm iteratively approximates the solution to an SLE by a linear combination of $\bW_t$-orthogonal direction vectors $\bp_j$, or \textit{conjugate directions}, such that

\noindent
\begin{equation}
    (\bp_t^i)^T \bW_t \bp_t^j = 0 \label{eq:CG_p_conjugate}
\end{equation}

\noindent
for any $i \neq j$. The algorithm involves two steps. First, it determines the new conjugate direction $\bp_t^i$. Second, it optimizes the magnitude $a_t^i$ of the step along the new direction $\bp_t^i$ with respect to a quadratic loss function $f(\bmu_A^{t,i}) = (\bmu_A^{t,i})^T \bW_t \bmu_A^{t,i} - \bz_t^T \bmu_A^{t,i}$, where $\bmu_A^{t,i} = \sum_{j=0}^i a_t^j \bp_t^j$ is the approximation of the $M$-dimensional SLE solution, $\bmu_A^t$, after $i$ iterations. If we define the error 

\noindent
\begin{equation}
    \be_t^i = \bmu_A^{t,i} - \bmu_A^t,
\end{equation}  

\noindent
then it can be shown \cite{CG_properties}, \cite{Painless_CG} that after each iteration $i$, the CG algorithm minimizes the error $E_t^i = (\be_t^i)^T \bW_t \be_t^i$ within the subspace $\be_0 + span\big\{ \bp_t^0, \bp_t^1, ..., \bp_t^i \big\}$. Therefore, after $M$ iterations the error $E_t^i$ converges to zero since $\bW_t$ is positive-definite of rank $M$. However, applying CG with $M$ iterations becomes as expensive as solving \eqref{eq:SLE} exactly. In this work we consider the performance CG-VAMP when only a few CG iterations $i$ are used to approximate $\bmu_A^t$ so that the resulting complexity of a CG-VAMP-based algorithm is close to AMP's.

\begin{algorithm}
\DontPrintSemicolon
\SetNoFillComment
\caption{CG for approximating $\bW \bmu = \bz$}

    \KwInitialization{ $\bmu = \bzero_M, \br^0 = \bp^0 = \bz, i = 0$}

    \While{$i < i_{max}$}
    {
        $\bd^i = \bW \bp^i$ \;
        $a^i = \frac{||\br^i ||^2}{ (\bp^i)^T \bd^i}$ \;
        $\bmu^{i+1} = \bmu^i + a^i \bp^i$ \;
        $\br^{i+1} = \br^i - a^i \bd^i$ \;
        $b^i = \frac{|| \br^{i+1} ||^2}{|| \br^{i} ||^2}$ \;
        $\bp^{i+1} = \br^{i+1} + b^i \bp^i = \bz - \bW \bmu^{i+1} + b^i \bp^i$ \;
        $i = i+1$
    }
    \KwOutput{$\bmu^{i+1}, a^i, b^i, ||\br^i||^2$}    

\end{algorithm}

\subsection{Large System Limit dynamics of CG-VAMP} \label{sec:CG_VAMP_LSL}

Using an approximated version of the LMMSE estimator instead of the exact one within CG-VAMP would additionally require redesigning the correcting scalar $\gamma_A^t$ from \eqref{eq:gamma_A_EP} and the SE update for $\vAB^t$. This problem was first considered in the work \cite{CG_EP}, where the authors showed that in the limit $N \rightarrow \infty$, the output of $\bmu_A^{t,i}$ of the CG algorithm with $i$ iterations almost surely converges to

\noindent
\begin{equation}
    \limN \bmu_A^{t,i} = \limN \sum_{j=0}^i r_t^{t,i}[j] \bPhi^j  \bz_t, \label{eq:zero_init_CG_LSL_model}
\end{equation}

\noindent
where $\bPhi = \bA \bA^T$, and $r_t^{t,i}[j]$ is a scalar function of $\vBA^t$ and $v_w$ only and its exact definition can be found in Appendix A. Then the authors applied \eqref{eq:zero_init_CG_LSL_model} to \eqref{eq:gamma_A_general} to obtain the solution for $\gamma_A^{t,i}$

\noindent
\begin{equation}
    \gamma_A^{t,i} = - \sum_{j=0}^i r_t^{t,i}[j] \chi^{j+1} \label{eq:zero_init_CG_gamma_A}
\end{equation}

\noindent
where $\chi_{j}$ represent the spectral moments

\noindent
\begin{equation}
    \chi_{j} = \inv{N} Tr \big\{ (\bS \bS)^j \big\} \label{eq:chi}
\end{equation}

\noindent
Similarly, by expanding the definition of the variance $\vAB^{t,i}$

\noindent
\begin{equation}
    \vAB^{t,i} \as \limN ||\xAB^t - \bx||^2 = \limN \Big|\Big|\bq_t - \frac{\bmu_A^{t,i}}{\gamma_A^{t,i}}\Big|\Big|^2 \label{eq:zero_init_CG_v_AB_general}
\end{equation}

\noindent
and using \eqref{eq:gamma_A_general_equality} and \eqref{eq:zero_init_CG_LSL_model}, the authors showed that

\noindent
\begin{align}
   \vAB^{t,i} \as (\gamma_A^i)^{-2} &\sum_{j = 0}^i \sum_{k = 0}^i \bigg( r_t^{t,i}[j] r_t^{t,i}[k]\big( \vBA^t \chi_{j+k+2} \nonumber\\
   &+ v_w \chi_{j+k+1} \big) \bigg) - \vBA^t \label{eq:zero_init_CG_v_AB}
\end{align}

While \eqref{eq:zero_init_CG_gamma_A} and \eqref{eq:zero_init_CG_v_AB} provide the means of theoretically studying the algorithm, these results are highly sensitive to the estimation error. To see this, first, we note that these results require the oracle access to the moments $\chi_j$, that can be estimated using black-box methods like in \cite{MC-divergence}. For that type of methods, the generated estimate of the moment $\chi_j$ has the variance proportional to $\chi_{2j}$ \cite{MC_matrix_moments_1}. Since we assume the normalization $\inv{N} Tr\{\bA \bA^T\} = \chi_1 = 1$, we have that 

\noindent
\begin{equation*}
    \chi_2 = \inv{N} Tr \big\{ (\bLambda - \chi_1)^2 \big\} + \chi_1 = \inv{N} Tr \big\{ (\bLambda - \chi_1)^2 \big\} + 1
\end{equation*}

\noindent
is greater than 1 if $\kappa(\bA) > 1$. Then, from the H\"older's inequality we have

\noindent
\begin{equation*}
    \chi_j \geq (\chi_2)^{\frac{j}{2}},
\end{equation*}

\noindent
which implies that the estimation variance $\chi_{2j}$ has exponential growth with $j$. As a result, the estimates $\Hat{\chi}_j$ of the spectral moments might be highly inaccurate for a larger number of CG iterations $i$. Moreover, these large spectral moments scale directly and indirectly (through $r_t^{t,i}[j]$) the variances $v_w$ and $\vBA^t$. In practice we do not have the direct access to them and can only use their estimated versions $\hvw$ and $\hvBA^t$ that are corrupted by some estimation error. In the estimators based on \eqref{eq:zero_init_CG_gamma_A} and \eqref{eq:zero_init_CG_v_AB} this estimation error is further magnified by exponentially in $j$ growing moments $\chi_j$. Since the SE \eqref{eq:zero_init_CG_v_AB} uses the moments $\chi_j$ of order up to $i^2 + 2$, even a small number $i$ might potentially lead to instability of the estimator for $\vAB^{t,i}$. In the simulation section we demonstrate that CG-VAMP equipped with the estimators based on \eqref{eq:zero_init_CG_gamma_A} and \eqref{eq:zero_init_CG_v_AB} exhibits unstable dynamics even with small $i$, which motives us to seek alternative asymptotic results for $\gamma_A^{t,i}$ and $\vAB^{t,i}$.

\section{Stable implementation of CG-VAMP} \label{sec:stable_estimation_in_CG_VAMP}

The first goal of the paper is to develop stable methods for estimating $\gamma_A^{t,i}$ and $\vAB^{t,i}$ without explicitly referring to the spectral moments $\chi_j$. We begin with the correction scalar $\gamma_A^{t,i}$, which must follow the general identity \eqref{eq:gamma_A_general_equality}

\noindent
\begin{equation}
    \limN \inv{N} \bq_t^T \bA^T \bmu_A^{t,i} \as \vBA^t \gamma_A^{t,i} \label{eq:gamma_A_equality}
\end{equation}

\noindent
Although this result cannot be used directly, it becomes helpful when we consider it together with the following result

\noindent
\begin{equation}
    \bz_t = \by - \bA \xBA^t = \bA \bx + \bw - \bA \bx - \bA \bq_t = \bw - \bA \bq_t \label{eq:z_t}
\end{equation}

\noindent
where we used \eqref{eq:y_measurements} and \eqref{eq:x_ba}. Thus we have

\noindent
\begin{equation}
    \limN \inv{N} \bw^T \bmu_A^{t,i} - \inv{N} \bz_t^T \bmu_A^{t,i} \as \vBA^t \gamma_A^{t,i}, \label{eq:z_mu_zero_init_CG}
\end{equation}

\noindent
which implies that we can recover $\gamma_A^{t,i}$ from

\noindent
\begin{equation}
    \gamma_A^{t,i} \as \limN \frac{\inv{N} \bw^T \bmu_A^{t,i} - \inv{N} \bz_t^T \bmu_A^{t,i}}{\vBA^t} \label{eq:zero_init_gamma_identity_impractical}
\end{equation}

\noindent
To complete this idea, we present a theorem that establishes the asymptotic dynamics of the last unknown component $\limN \inv{N} \bw^T \bmu_A^{t,i}$ in the context of CG-VAMP.

\begin{theorem} \label{theorem:zero_init_gamma_estimator}
    Let $\bmu_A^{t,i}$ be the output of the zero-initialized CG after $i$ inner-loop iterations. For $j=0,...,i$, let $a_t^j$ and $b_t^j$ be as in Algorithm 2. Define a sequence of scalar functions $\eta_t^{j}$ as
    
    \noindent
    \begin{equation}
        \eta_t^{j+1} = v_w \Big( \delta - \inv{N} \bz_t^T \bmu_t^{j+1} \Big) + b_t^{j} \eta_t^{j} \label{eq:LSL_gamma_A_rec_1}
    \end{equation}

    \noindent
    with $\eta_t^0 = \delta v_w$. Then, under Assumptions 1-3, we have that
    
    \noindent
    \begin{equation}
        \limN \inv{N} \bw^T \bmu_A^{t,i} \as \limN \sum_{j=0}^i a_t^j \eta_t^j
    \end{equation}
    
\end{theorem}

\begin{proof}
    See Appendix D.
\end{proof}

With this theorem, one can define an estimator for $\gamma_A^{t,i}$ as

\noindent
\begin{equation}
    \hgam_A^{t,i} = \frac{\sum_{j=0}^i a_t^j \heta_t^j - \inv{N} \bz_t^T \bmu_A^{t,i}}{\hvBA^t} \label{eq:iterative_gamma_zero_init_CG}
\end{equation}

\noindent
where $\heta_t^j$ is computed from \eqref{eq:LSL_gamma_A_rec_1} with $v_w$ replaced by its estimated value. Importantly, \eqref{eq:iterative_gamma_zero_init_CG} does not rely on the knowledge of $\chi_j$ contrary to \eqref{eq:zero_init_CG_gamma_A} and can be naturally implemented in the body of CG to update the correction scalar after each iteration. Additionally, the computational cost of \eqref{eq:iterative_gamma_zero_init_CG} is dominated by one inner-product of $M$-dimensional vectors, which makes the cost of the estimator negligible compared to the cost of the CG routine. Lastly, as will be shown in the simulation section, the range of $i$ for which the estimator\eqref{eq:iterative_gamma_zero_init_CG} demonstrates stable and accurate performance is by an order larger than for the estimator based on \eqref{eq:zero_init_CG_gamma_A}.

\subsection{Practical estimation of $\vAB^{t,i}$ in CG-VAMP}

Next, we propose an asymptotically consistent estimator for the variance $\vAB^{t,i}$ that can be naturally implemented within CG and has negligible computational and memory costs. For this, we expand the result \eqref{eq:zero_init_CG_v_AB_general} and use the identity \eqref{eq:gamma_A_general_equality} to obtain

\noindent
\begin{align}
    &\vAB^{t,i} \as \limN \inv{N} ||\bq_t - (\gamma_A^{t,i})^{-1} \bA^T \bmu_A^{t,i}||^2 \nonumber\\
    &\as (\gamma_A^{t,i})^{-2} \limN \inv{N} ||\bA^T \bmu_A^{t,i}||^2 - \vBA^t \label{eq:v_AB_practical_but_slow}
\end{align}

\noindent
Although this result can be used directly by substituting in the estimated values $\hgam_A^{t,i}$ and $\hvBA^t$, executing it after each CG iteration $i$ would double the cost of Block A due to the additional matrix-vector product $\bA^T \bmu_A^{t,i}$. Therefore, next we reformulate \eqref{eq:v_AB_practical_but_slow} to obtain an estimator of $\vAB^{t,i}$ that utilizes the results computed in the the CG routine to avoid additional matrix-vector products.

\begin{theorem} \label{theorem:zero_init_CG_v_ab_efficient}
    Let $\bmu_A^{t,i}$ be the output of the CG algorithm from Algorithm 2 and $\gamma_A^{t,i}$ satisfy \eqref{eq:iterative_gamma_zero_init_CG}. Then, under Assumptions 1-3, we have
    
    \noindent
    \begin{equation}
        \vAB^{t,i} \overset{a.s.}{=} \lim_{N \rightarrow \infty}  \frac{ \zeta_t^{i} - v_w \frac{|| \bmu_A^{i,t} ||^2}{N} }{ \vBA^t \big(\gamma_A^{t,i}\big)^2} - \vBA^t
        \label{eq:th:v_ab_efficient}
    \end{equation}
    
    \noindent
    where the scalar $\zeta_t^{i}$ is iteratively defined as
    
    \noindent
    \begin{equation}
        \zeta_t^{i} = \zeta_t^{i-1} + a_t^i  \inv{N} ||\br_t^i||^2 \label{eq:CG_zeta}
    \end{equation}
    
    \noindent
    with $\zeta_t^{0} = 0$ and $\br_t^i$ and $a_t^i$ are as in Algorithm 2.

\end{theorem}

\begin{proof}
    See Appendix E.
\end{proof}

\noindent
Thus we propose the following estimator of $\vAB^{t,i}$

\noindent
\begin{equation}
    \hvAB^{t,i} = \inv{N}  \frac{ \zeta_t^{i} - \hvw || \bmu_A^{t,i} ||^2 }{\hvBA^t \big( \hgam_A^{t,i} \big)^{2}} - \hvBA^t
    \label{eq:v_ab_efficient}
\end{equation}

\noindent
where instead of the true values of $\gamma_A^{t,i}$, $v_w$ and $\vBA^t$ we used their estimated versions. The estimator \eqref{eq:v_ab_efficient} can be efficiently implemented in the body of CG and computed at each iteration at the additional cost of $O(M)$ provided an estimate of $\gamma_A^{t,i}$.

\section{Adaptive CG in CG-VAMP} \label{sec:ACG}

The developed tools in the previous section allow one to efficiently estimate the correction scalar $\gamma_A^{t,i}$ and the SE for $\vAB^{t,i}$ after each CG iteration. Yet, there is still the question of how many of those CG iterations we should do at each iteration $t$ in order to obtain consistent progression and efficient performance of CG-VAMP. Note that within the family of Message Passing algorithms with a 1D SE it was shown \cite{VAMP}, \cite{OAMP} that the LMMSE estimator is the optimal choice with respect to the SE for Block A so that

\noindent
\begin{equation}
    \vAB^{t,i} \geq \vAB^{t,M} \label{eq:SE_inequality_1}
\end{equation}

\noindent
for $i \leq M$. The first consequence of this inequality is that CG-VAMP has a lower convergence rate (in terms of outer-loop iterations $t$) in comparison to VAMP and this is usually a reasonable compromise. However, this inequality also means CG-VAMP may converge to a spurious fixed point that is far from the VAMP fixed point. Moreover, if the approximation produced by CG is sufficiently bad, then $\vAB^{t,i}$ might even increase with respect to $\vAB^{t-1}$. Based on numerical experiments in the current and the previous work \cite{OurPaper}, we have observed that the number of the inner-loop iterations sufficient to achieve a certain reduction of $\vAB^{t,i}$ per-outer-loop iteration significantly changes with $t$ and therefore motivates the benefit in the adaptive choice of $i[t]$ at each $t$.

To achieve a monotonic reduction of the variance $\vAB^{t,i}$ after each outer-loop iteration $t$, while keeping $i$ as small as possible, we can iterate the CG algorithm until the following inequality is met

\noindent
\begin{equation}
    \hvAB^{t,i} < c \hvAB^{t-1,i[t-1]} \label{eq:CG_stopping_rule}
\end{equation}

\noindent
where the constant $c<1$ defines the expected per-outer-loop iteration reduction of the variance $\vAB^{t,i}$. Since the reduction of the intrinsic variance $\hvAB^{t,i}$ depends on the performance of both blocks, Block A and Block B, choosing the right scalar $c$ is critical for balancing the workload between the blocks. On one hand, we observe that the relative improvement of $\vAB^{t,i}$ after every next CG iteration $i$ decreases and eventually becomes negligible. Thus, to avoid inefficient operation of Block A, we might want to keep $c$ relatively large and terminate CG earlier. On the other hand, the first CG iterations provide the most reduction in term of the SE and terminating CG too early implies that we put more work on the denoising block. To balance the workload between the blocks, we suggest to compare the relative expected improvement of $\vAB^{t,i}$ after each iteration $i$ and proceed to \eqref{eq:CG_stopping_rule} when the following inequality is met

\noindent
\begin{equation}
    \frac{\hvAB^{t,i-1} - \hvAB^{t,i}}{\hvAB^{t,i}} < \Delta \label{eq:CG_stopping_rule_2}
\end{equation}

\noindent
where $\Delta$ is some positive scalar. This inequality suggests that if the last inner-loop iteration was efficient enough, then we should continue iterating the CG algorithm regardless whether we have already met the inequality \eqref{eq:CG_stopping_rule}. The choice of the constant $\Delta$ depends on the parameters of the measurement system \eqref{eq:y_measurements} and the chosen denoiser, and should be adapted for different applications individually. Lastly, we bound $i$ by some $i_{max}$ in the case if we chose too ambitious reduction level $c$, which is not achievable even for VAMP or within a reasonable number of CG iterations.

These stopping criteria together with the developed tools from the previous section are used to construct an \textit{Adaptive CG (ACG)} shown on Algorithm 3. The first block of the algorithm implements the function $CG$ that represents the $i$-th iteration steps of CG from Algorithm 2 and outputs the vector $\bmu_A^{t,i}$ and the data required for estimating $\gamma_A^{t,i}$ and $\vAB^{t,i}$. The memory requirements and the computational time of ACG are dominated by executing the CG routing where we need to compute the matrix-vector product $\bd_t^i = \bW_t \bp_t^i$. Such an operation in general is of complexity $O(M^2)$, but can be significantly accelerated when the measurement matrix $\bA$ has a fast implementation where the complexity can reduce down to $O(N \log_2 N)$ or even smaller. The performance of the CG-VAMP algorithm with ACG in recovering large scale natural images measured by a fast operator will be demonstrated in the simulation section.

\begin{algorithm}
\DontPrintSemicolon
\SetNoFillComment
\caption{Adaptive Conjugate Gradient}

    \KwInitialization{ $\bmu_A^{0,t} = \bm{0}, \br_t^0 = \bp_t^0 = \bz_t, \nu_t^0 = 0, \eta_t^0 = \delta \hvw, \hzeta_t^0 = 0, \hvAB^{t,0} = \infty, i = 0$}

    \While{$i < i_{max} \: and \: \Big( \Delta(i) > \Delta \: or \: \hvAB^{t,i} > c \hvAB^{t-1} \Big)$}
    {
    
        \nonl \KwBlockCGRoutine{
            $\big(\bmu_A^{t,i+1}, a_t^i, b_t^i, ||\br_t^i||^2\big) = CG(i, \bmu_A^{t,0}, \bp_t^0)$ \;
        }
        \nonl \KwBlockCGDivergence{
            $\heta_t^{i+1} = \hvw \Big( \delta - \inv{N} \bz_t^T \bmu_A^{t,i+1} \Big) + b_t^{i} \heta_t^{i}$ \;
            $\hgam_A^{t,i+1} \as \frac{\sum_{j=0}^{i+1} a_t^j \heta_t^j - \inv{N} \bz_t^T \bmu_A^{t,i+1}}{\hvBA^t}$ \;
        }
        
        \nonl \KwBlockCGMSE{
            $\zeta_t^{i+1} = \zeta_t^{i} + a_t^i  \inv{N} ||\br_t^i||^2$ \;
            $\hvAB^{t,i+1} = \frac{ \zeta_t^{i+1} - \hvw \frac{|| \bmu_A^{i,t} ||^2}{N} }{\hvBA^t \big( \hgam_t^{i+1} \big)^{2}} - \hvBA^t$
        }
        
        $\Delta(i) = \frac{\hvAB^{t,i} - \hvAB^{t,i+1}}{\hvAB^{t,i+1}}$\;
        
        $i = i+1$ \;
    }
    \KwOutput{$\bmu_A^{t,i+1}, \hvAB^{t,i+1}, \hgam_A^{t,i+1}$}

\end{algorithm}

\section{Warm-starting the CG algorithm in CG-VAMP} \label{sec:warm_starting}

In this section we discuss an acceleration method that is common for certain optimization methods such as Sequential Quadratic Programming \cite{Opt_ref1} and Interior-Point methods \cite{Opt_ref2} -- warm-starting. In the CG-VAMP algorithm, at each outer-loop iteration we have to approximate a solution to the updated SLE $\bW_t \bmu_t = \bz_t$ and a natural question arises: how different are these SLE at two consecutive iterations? If the pair of the SLE are close, then due to asymptotic linearity of the CG algorithm from \eqref{eq:zero_init_CG_LSL_model}, the outputs $\bmu_A^{t,i}$ and $\bmu_A^{t-1,i} $ should be close as well. Thus it is natural to consider initializing the CG algorithm with "an approximation" of its expected \textit{state} - the state from the previous outer-loop iteration. In particular, we propose to initialize the starting CG approximation $\bmu_A^{t,0}$ as

\noindent
\begin{equation}
    \bmu_A^{t,0} = \bmu_A^{t-1,i} \label{eq:mu_init}
\end{equation}

\noindent
and the memory term, $\bp_t^i$, of CG as 

\noindent
\begin{equation}
    \bp_t^0 = \bz_t - \bW_t \bmu_A^{t-1,i} + b_{t-1}^{i-1} \bp_{t-1}^{i-1} \label{eq:p_init}
\end{equation}

\noindent
Here, we do not use the initialization $\bp_t^0 = \bp_{t-1}^i$ because the residual term $\br_t^0 = \bz_t - \bW_t \bmu_A^{t-1,i} $ is not the same as $\br_{t-1}^i$ and using \eqref{eq:p_init} accounts for the difference in the SLE at $t$ and $t-1$.

The idea of using a warm-starting strategy is supported by the similarity of the SLE \eqref{eq:SLE} at two consecutive iterations given the algorithm has not progressed substantially. To see this, first, we note that in the SLE \eqref{eq:SLE}, the matrix $\bW_t$

\noindent
\begin{equation*}
    \bW_t = v_w \bI_M + \vBA^t \bA \bA^T = \bU \big( v_w \bI_M + \vBA^t \bLambda \big) \bU^T
\end{equation*}

\noindent
changes only in the eigenvalues, while the eigenvectors of $\bW_t$ remain the same for all $t$. Therefore, if $\vBA^t$ and $\vBA^{t-1}$ are close, then $\bW_t$ will not change much from the previous iteration. 

Second, we observe that when the intrinsic variance $\vBA^t$ does not change much at two consecutive iterations, the corresponding error vector $\bq_t$ and $\bq_{t-1}$ are similar as well. As a result, the constant term $\bz_t = \bw - \bA \bq_t$ remains close to $\bz_{t-1}$. That, together with the similarity of $\bW_t$ at two consecutive iterations, provides the motivation for reusing the CG information generated at the previous outer-loop iterations.

\subsection{Rigorous dynamics of Block A in WS-CG-VAMP} \label{subsec:warm-starting_appropriate_correction}

As presented in Section \ref{sec:CG_VAMP_LSL}, when the CG algorithm is initialized with $\bmu_A^{t,0} = \bzero_M$, the asymptotic dynamics of the resulting CG-EP algorithm can be defined through a 1D State Evolution (SE). However, when we use the initializations \eqref{eq:mu_init}-\eqref{eq:p_init} and keep the same structure of $\xAB^t$ from Algorithm 1 with a single correction scalar $\gamma_A^{t,i}$, the resulting CG-VAMP algorithm loses the SE property. To see this, first, we can sequentially apply the initializations \eqref{eq:mu_init}-\eqref{eq:p_init} to show that

\noindent
\begin{equation}
    \bmu_A^{t,i} = \bg_A^{i} \big( \bz_t, \big\{ \bmu_A^{\tau,i} \big\}_{\tau = 0}^{t-1}, \big\{ \bp_{\tau}^{i-1} \big\}_{\tau = 0}^{t-1} \big)
\end{equation}

\noindent
Next, we note that both $\bmu_A^{\tau,i}$ and $\bp_{\tau}^{i-1}$ depend on $\bz_{\tau}$, which depends on $\bw$ and $\bq_{\tau}$ as we know from \eqref{eq:z_t}. Therefore we have that the WS-CG output $\bmu_A^{t,i} = \bg_A^i\big( \bQ_t, \bw \big)$, where $\bQ_t = \big( \bq_t, \bq_{t-1}, ..., \bq_{1}, \bq_{0} \big)$ is a function of the whole history of the error vectors $\bq_{\tau}, \tau = 0,...,t$ contrary to the zero-initialized CG, where $\bmu_A^{t,i}$ is a function of only $\bq_t$ as seen from \eqref{eq:zero_init_CG_LSL_model}. However, as shown in \cite{UnifiedSE}, one needs to construct a correction scalar $\gamma_A^{t,\tau}$ for each error vector $\bq_{\tau}$ the function $\bg_A$ depends on. Thus, for WS-CG the new form of the update $\xAB^t$ must be changed to \cite{UnifiedSE}

\noindent
\begin{equation}
    \xAB^t = \frac{1}{\sum_{\tau = 0}^t \gamma_A^{t,\tau,i}} \Big( \sum_{\tau = 0}^t \gamma_A^{t,\tau, i} \xBA^{\tau} - \bA^T \bmu_A^{t,i} \Big) \label{eq:x_ab_warm_started}
\end{equation}

\noindent
in order to preserve the asymptotic properties \eqref{eq:h_t_gaussian} and \eqref{eq:h_q_independence}, and be able to define the SE of the form $\vAB^{t,i} \as \bff_A^i(\bPsi_t)$, where $(\bPsi)_{\tau,\tau'} \as \limN \inv{N} \bq_{\tau}^T\bq_{\tau'}$. Here the correction scalars $\gamma_A^{t,\tau, i}$ are the generalized versions of \eqref{eq:gamma_A_general} for each error vector $\bq_{\tau}$ individually 

\noindent
\begin{equation}
    \gamma_A^{t,\tau, i} \as \lim_{N \rightarrow \infty} \inv{N} \nabla_{\bq_{\tau}} \cdot \Big( \bA^T \bg_A^i\big( \bQ_t, \bw \big) \Big) \label{eq:WS_gamma_A_assymptotic}
\end{equation}

\noindent
and must follow the identity

\noindent
\begin{equation}
    \limN \inv{N} \bq_{\tau}^T \bA^T \bmu_A^{t,i} \as \sum_{\tau = 0}^t \psi_{t,\tau} \gamma_A^{t,\tau,i} \label{eq:WS_gamma_A_equality}
\end{equation}

\noindent
which is a generalized version of the identity \eqref{eq:gamma_A_equality}. 

As a result of changing the structure of $\xAB^t$, the SE of Block A changes as well. Let $C_A = \frac{1}{\sum_{\tau = 0}^t \gamma_A^{t,\tau,i}}$ and $\psi_{\tau, \tau'} =(\bPsi)_{\tau,\tau'}$. Then, we can show that

\noindent
\begin{align}
    &\vAB^{t,i} \as \lim_{N \rightarrow \infty} \inv{N} ||\xAB^t - \bx||^2 \nonumber\\
    &\overset{(a)}{=} \lim_{N \rightarrow \infty} \frac{C_A^2}{N} \bigg| \bigg| \sum_{\tau = 0}^t \gamma_A^{t,\tau,i} \bq_{\tau} - \bA^T \bmu_A^{t,i} \bigg| \bigg|^2 \nonumber\\
    &= \lim_{N \rightarrow \infty} \frac{C_A^2}{N} \Big( \big| \big| \sum_{\tau = 0}^t \gamma_A^{t,\tau,i} \bq_{\tau} \big| \big|^2 - 2\sum_{\tau = 0}^t \gamma_A^{t,\tau,i} \bq_{\tau}^T \bA^T \bmu_A^{t,i} \nonumber\\
    &+\big| \big| \bA^T \bmu_A^{t,i} \big| \big|^2 \Big) \nonumber\\
    &\as \lim_{N \rightarrow \infty} C_A^2 \Big( \frac{\big| \big|\bA^T \bmu_A^{t,i} \big| \big|^2}{N} - \sum_{\tau, \tau' = 0}^t \psi_{\tau,\tau'} \gamma_A^{t,\tau,i} \gamma_A^{t,\tau',i} \Big) \label{eq:v_ab_WS_start}
\end{align}

\noindent
where in (a) we used \eqref{eq:x_ab_warm_started} and the last step of \eqref{eq:v_ab_WS_start} is due to \eqref{eq:WS_gamma_A_equality}. Now, in order to derive the closed-form solutions to \eqref{eq:WS_gamma_A_assymptotic} and \eqref{eq:v_ab_WS_start}, we follow a similar strategy as in Section \ref{sec:CG_VAMP_LSL} and first show that in the limit $N \rightarrow \infty$ WS-CG corresponds to a linear mapping of all $\bz_{\tau}$ for $\tau = 0,...,t$.

\begin{theorem} \label{theorem:bz_tau_to_mu}
    
    Let $\bmu_A^{t,i}$ be the output of WS-CG with the initializations \eqref{eq:mu_init} and \eqref{eq:p_init}. Then, under Assumptions 1-3, we have that
    
    \noindent
    \begin{equation}
        \bmu_A^{t,i} = \sum_{\tau = 0}^{t}  \sum_{k = 0}^{(t - \tau) i} r_{\tau}^{t,i}[k] \bPhi^k   \bz_{\tau} \label{eq:bz_tau_to_bmu}\\
    \end{equation}
    
    \noindent
    where $r_{\tau}^{t,i}[k]$ is a scalar function of $\delta$, $v_w$ and $\big\{\vAB^{\tau'}\big\}_{\tau'=\tau}^t$ and is defined in Appendix A.
    
\end{theorem}

\begin{proof}
    See Appendix A.
\end{proof}

As seen from the theorem, WS-CG has a similar asymptotic structure to \eqref{eq:zero_init_CG_LSL_model} for zero-initialized CG and, in fact, as shown in Appendix A, the dependence of $\bmu_A^{t,i}$ from \eqref{eq:bz_tau_to_bmu} on $\bz_t$ is exactly the same as in \eqref{eq:zero_init_CG_LSL_model}. 

Next, by substituting \eqref{eq:bz_tau_to_bmu} into \eqref{eq:WS_gamma_A_assymptotic} and \eqref{eq:v_ab_WS_start}, we can obtain the closed-form solutions for $\gamma_A^{t,\tau,i}$ and $\vAB^{t,i}$ respectively. The results are formulated in the following theorem

\begin{theorem} \label{theorem:WS_CG_correction_and_SE}
    
    Let $\bmu_A^{t,i}$ be the output of the WS-CG algorithm with the initializations \eqref{eq:mu_init} and \eqref{eq:p_init}, and $r_{\tau}^{t,i}[k]$ be as in \eqref{eq:bz_tau_to_bmu}. Then, under Assumptions 1-3, the scalar $\gamma_A^{t,\tau,i}$ from \eqref{eq:WS_gamma_A_assymptotic} almost surely converges to 
     
    \noindent
    \begin{equation}
        \gamma_A^{t,\tau,i} \overset{a.s.}{=} - \sum_{k = 0}^{(t - \tau) i} r_{\tau}^{t,i}[k] \chi_{k+1} \label{eq:WS_CG_correction_scalars}
    \end{equation}
    
    \noindent
    where $\chi_j$ is as in \eqref{eq:chi}. Additionally, the variance $\vAB^{t,i}$ from \eqref{eq:v_ab_WS_start} evolves as
    
    \noindent
    \begin{equation}
        \vAB^{t,i} \overset{a.s.}{=} \frac{\Omega_t^i - \sum_{\tau, \tau' = 0}^t \gamma_A^{t,\tau,i} \gamma_A^{t,\tau',i} \psi_{\tau,\tau'}}{\Big(\sum_{\tau = 0}^t \gamma_A^{t,\tau,i}\Big)^2}  \label{eq:v_ab_WS_CG}
    \end{equation}
    
    \noindent
    with
    
    \noindent
    \begin{align}
        \Omega_t^i = \sum_{\tau,\tau' = 0}^{t} \sum_{j = 0}^{(t - \tau) i} &\sum_{k = 0}^{(t - \tau') i} r_{\tau}^{t,i}[j] r_{\tau'}^{t,i}[k] \big(\psi_{\tau,\tau'} \chi_{j+k+2} \nonumber\\
        &+ v_w \chi_{j+k+1} \big) \label{eq:Omege}
    \end{align}

\end{theorem}

\begin{proof}
    See Appendix B.
\end{proof}

As we see from Theorem \ref{theorem:WS_CG_correction_and_SE}, the asymptotic results for the correction scalars and for the variance $\vAB^{t,i}$ are structurally similar for both WS-CG and the zero-initialized CG. Unfortunately, the practical drawbacks of the two pairs of results are the same as well and, moreover, worse for WS-CG case. In particular, \eqref{eq:WS_CG_correction_scalars} relies on the access to the oracle spectral moments $\chi_j$ of order at least $t i$ for $\gamma_A^{t,i}$ contrary to the zero-initialized CG case, where the maximum moment order was $i$. An even worse situation with respect to the moment order is observed in \eqref{eq:v_ab_WS_CG}, which also requires the access to the cross-correlation measures $\psi_{\tau,\tau'}$ and is harder to estimate than just $\vBA^t$ that is the sufficient statistics for \eqref{eq:zero_init_CG_gamma_A}. In this work, we estimate $\psi_{\tau,\tau'}$ based on the following asymptotic result

\noindent
\begin{align}
    &\lim_{N \rightarrow \infty} \inv{N} \bz_{\tau}^T \bz_{\tau'} - \delta v_w \nonumber\\
    &= \lim_{N \rightarrow \infty} \inv{N} (\bw - \bA \bq_{\tau})^T (\bw - \bA \bq_{\tau'}) - \delta v_w \nonumber\\
    &\overset{a.s.}{=} \lim_{N \rightarrow \infty} \inv{N} \bq_{\tau}^T \bA^T \bA \bq_{\tau'} \overset{a.s.}{=} \lim_{N \rightarrow \infty} \inv{N} \bq_{\tau}^T \bq_{\tau'} = \psi_{\tau,\tau'} \label{eq:psi_LSL_estimator}
\end{align}

\noindent
where we used the following two asymptotic properties 

\noindent
\begin{gather}
    \limN \inv{N} \bw^T \bA \bq_t \as 0 \label{eq:w_Aq_independence} \\
    \limN \inv{N} \bq_{\tau}^T \bA^T \bA \bq_{\tau'} \as \limN \inv{N} \bq_{\tau}^T \bq_{\tau'} \label{eq:Aq_norm}
\end{gather}

\noindent
proved in \cite{UnifiedSE} to hold in Message Passing algorithms with multidimensional SE under Assumptions 1-3. In the simulation section, we confirm the validity of the SE \eqref{eq:v_ab_WS_CG} provided the oracle values of $\chi_j$, $v_w$ and $\psi_{\tau,\tau'}$, but also demonstrate the instability of WS-CG-VAMP when \eqref{eq:WS_CG_correction_scalars} and \eqref{eq:v_ab_WS_CG} are used to construct the corresponding estimators without the oracle information. Therefore, although Theorem \ref{theorem:WS_CG_correction_and_SE} provides theoretical tools for studying the asymptotics of WS-CG-VAMP, we need an alternative robust method for estimating $\gamma_A^{t,\tau,i}$ and $\vAB^{t,i}$ in practical scenarios.

\subsection{Robust estimation of Block A parameters in WS-CG-VAMP} \label{section:WS_CG_iterative_gamma}

We begin with deriving a more robust method of estimating $\vAB^{t,i}$ without explicitly referring to $\chi_j$ within WS-CG-VAMP. It turns out that by leveraging one of the main properties \eqref{eq:h_q_independence} of VAMP-like Message Passing algorithms, we can obtain the following result

\noindent
\begin{align}
    &\lim_{N \rightarrow \infty} \inv{N} ||\xAB^t - \xBA^t||^2 - \vBA^t \nonumber\\
    &= \lim_{N \rightarrow \infty} \inv{N} ||\bx + \bh_t - \bx - \bq_t||^2 - \vBA^t \nonumber\\
    &\overset{a.s.}{=} \vAB^t + \vBA^t - \vBA^t = \vAB^t
\end{align}

\noindent
Importantly, this result is invariant to the structure of $\xAB^t$ and holds for both CG-VAMP and WS-CG-VAMP. Thus, by subsisting $\hvBA^t$ into the last result gives us an estimator for $\vAB^t$

\noindent
\begin{equation}
    \hvAB^t = \inv{N} ||\xAB^t - \xBA^t||^2 - \hvBA^t \label{eq:v_ab_general_estimator}
\end{equation}

Next, to derive an alternative method for estimating $\gamma_A^{t,\tau,i}$, we follow the same idea as in Section \ref{sec:stable_estimation_in_CG_VAMP} where we derived an asymptotic iterative identity for $\gamma_A^{t,i}$ for zero-initialized CG. In the case of WS-CG, the key identity \eqref{eq:z_mu_zero_init_CG} takes a more general form as the consequence of \eqref{eq:WS_gamma_A_equality}

\noindent
\begin{equation}
    \limN \inv{N} \bw^T \bmu_A^{t,i} - \inv{N} \bz_t^T \bmu_A^{t,i} \as \sum_{\tau = 0}^t \psi_{t,\tau} \gamma_A^{t,\tau,i} \label{eq:z_mu_WS_CG}
\end{equation}

\noindent
Then, by defining a matrix $\bZ_t = (\bz_0, \bz_1, ..., \bz_t)$, the set of scalars $(\bgamma_A^{t,i})_{\tau} = \gamma_A^{t,\tau,i}$ for $\tau = 0,...,t$ can be recovered from

\noindent
\begin{equation}
    \bgamma_A^{t,i} \as \lim_{N \rightarrow \infty} \bPsi_t^{-1} \big( \inv{N} \bw^T \bmu_A^{t,i} \bone_t - \inv{N} \bZ_t^T \bmu_A^{t,i} \big) \label{eq:Gamma_t_i_WS_CG}
\end{equation}

\noindent
where the invertibility of $\bPsi_t$ was confirmed in \cite{CG_EP}. The following theorem completes this idea and presents an iterative process that defines the asymptotic behaviour of $\gamma_A^{t,\tau,i}$ within WS-CG-VAMP.

\begin{theorem} \label{theorem:alternative_gamma_t_tau_estimator}
    
    Let $\bmu_A^{t,i}$ be the output of the WS-CG algorithm with initializations \eqref{eq:mu_init} and \eqref{eq:p_init} after $i$ inner-loop iterations. For $j=0,...,i$, let $a_t^j$ and $b_t^j$ be the scalars computed as in the CG algorithm. Define a recursion
    
    \noindent
    \begin{gather}
        \nu_t^i = \nu_t^{i-1} + a_t^{i-1} \eta_t^{i-1} \\
        \bsigma_t^i = \bPsi_t^{-1} \big( \nu_t^i \bone_t - \inv{N} \bZ_t^T \bmu_A^{t,i}\big) \\
        \eta_t^i = v_w \big(\delta - \nu_t^i + \vBA^t ||\bgamma_A^{t,i}||_1\big) + b_t^{i-1} \eta_t^{i-1}
    \end{gather}

    \noindent
    with the initializations $\nu_0^0 = 0$, $\eta_0^0 = \delta v_w$ for $t=0$, and $\nu_t^0 = \nu_{t-1}^i$, $\eta_t^0 = v_w \big(\delta - \nu_{t-1}^i + \vBA^t ||\bgamma_A^{t-1,i}||_1 \big) + b_{t-1}^{i-1} \eta_{t-1}^{i-1}$ for $t>0$. Then, under Assumptions 1-3, we have that
    
    \noindent
    \begin{equation}
        \bgamma_A^{t,i} \overset{a.s.}{=} \limN \bsigma_t^{i} \label{eq:iterative_gamma_WS_CG}
    \end{equation}

\end{theorem}

\begin{proof}
    See Appendix D.
\end{proof}

From the practical point of view, estimating the whole set  $\bgamma_A^{t,\tau,i}$ requires an access to an estimate of $\bPsi_t$ that we can compute from \eqref{eq:psi_LSL_estimator} with $v_w$ replaced by $\hvw$. In the simulation section, we demonstrate that WS-CG-VAMP equipped with the variance $\vAB^{t,i}$ estimator \eqref{eq:v_ab_general_estimator} and the estimator of $\gamma_A^{t,\tau,i}$ based on Theorem \ref{theorem:alternative_gamma_t_tau_estimator} exhibits stable dynamics for a much larger range of $i$ and $t$ compared to the same algorithm but where $\gamma_A^{t,\tau,i}$ is estimated based on \eqref{eq:WS_CG_correction_scalars}. Yet, when the algorithm is progressing and the variance $\vBA^t$ is decreasing, the matrix $\bPsi_t$ becomes ill-conditioned (since $(\bPsi_t)_{\tau,\tau} = \vBA^{\tau}$) and a small error in estimating $\psi_{\tau,\tau'}$ leads to a large error in estimating the inverse $\bPsi_t^{-1}$. From the numerical study, we have observed that for a large number of outer-loop iterations $t$, the WS-CG-VAMP with \eqref{eq:psi_LSL_estimator} used to estimate $\psi_{\tau',\tau}$ might eventually diverge. To the best of our knowledge, there is no method that demonstrates higher robustness in estimating $\psi_{\tau,\tau'}$ and instead we seek a more basic structure of $\xAB^t$ that involves a simpler correction of WS-CG.

\subsection{Approximate update $\xAB^t$ in WS-CG-VAMP} \label{subsec:single_correction_WS_CG_VAMP}

In the previous subsections we considered using the update \eqref{eq:x_ab_warm_started} for $\xAB^t$ with the full set of correction scalars $\gamma_A^{t,\tau,i}$ for WS-CG-VAMP. However, from the numerical experiments we have found out that for small $t$ the magnitude of the correction terms $O_t^{\tau} = \gamma_A^{t,\tau,i} \xBA^{\tau}$ in \eqref{eq:x_ab_warm_started} rapidly decays as $\tau$ decreases. This suggests that \eqref{eq:x_ab_warm_started} might be well-approximated by the update

\noindent
\begin{align*}
    \xAB^t &= (\gamma_A^{t,t,i})^{-1} \big( \gamma_A^{t,t,i} \xBA^t - \bA^T \bmu_A^{t,i} \big) \nonumber\\
    &= \xBA^t - (\gamma_A^{t,t,i})^{-1} \bA^T \bmu_A^{t,i}
\end{align*}

\noindent
which is exactly the same as the update for $\xAB^t$ in regular CG-VAMP since $\gamma_A^{t,t,i} = \gamma_A^{t,i}$. Therefore, an alternative, although non rigorous, approach to implement WS-CG-VAMP is to keep the same update rule of $\xAB^t$ as in CG-VAMP and use \eqref{eq:iterative_gamma_zero_init_CG} to estimate $\gamma_A^{t,i}$. While it is expected that such a message passing algorithm would eventually diverge due to accumulation of the correlation between $\bh_t$ and $\bq_t$, we will numerically demonstrate that such a WS-CG-VAMP algorithm is stable for many outer-loop iterations and can substantially outperform the standard CG-VAMP algorithm.

\section{Numerical experiments} \label{sec:numerical_experiments}

In this section we numerically compare the performance of CG-VAMP and WS-CG-VAMP with different correction methods. In the experiments we consider the measurement model \eqref{eq:y_measurements} where $\bA$ corresponds to the Fast ill-conditioned Johnson-Lindenstrauss Transform (FIJL)\cite{NS-VAMP}. In our case, the FIJL operator $\bA = \bS \bP \bH \bD$ is composed of the following matrices \cite{NS-VAMP}: the values of the diagonal matrix $\bD$ are either $-1$ or $1$ with equal probability; the matrix $\bH$ is some fast orthogonal transform. In our simulations, we chose $\bH$ to be the Discreet Cosine Transform (DCT); The matrix $\bP$ is a random permutation matrix and the matrix $\bS$ is an $M$ by $N$ matrix of zeros except for the main diagonal, where the singular values follow geometric progression leading to the desired condition number \cite{VAMP}. 

Although the FIJL operator is rather artificial, it is convenient for evaluating the performance of algorithms since it acts as a prototypical ill-conditioned CS matrix, requires no storing of matrices and has a fast implementation. Additionally, the FIJL operator that we consider enables us to directly implement VAMP for comparison in our experiments, since 

\noindent
\begin{equation}
    \bA \bA^T = \bS \bP \bH \bD \bD^T \bH^T \bP^T \bS^T = \bS \bS^T
\end{equation}

\noindent
and, therefore, the matrix inverse

\noindent
\begin{equation*}
    \bW_t^{-1} = (v_w \bI_M + \vBA^t \bA \bA^T)^{-1} = (v_w \bI_M + \vBA^t \bS \bS^T)^{-1}
\end{equation*}

\noindent
requires inverting only a diagonal matrix. However, here we emphasize that the CG-VAMP and WS-CG-VAMP algorithms that we implement do not utilize the fact that $\bW_t$ is diagonal and operate as if $\bW$ is an arbitrary matrix.

In the following experiments, for all the algorithms we design the same Block B, where we use a Denoising CNN\footnote{In this work we use the DnCNN from the D-AMP Toolbox available on \url{https://github.com/ricedsp/D-AMP_Toolbox}} (DnCNN) with $20$ layers and the Black-Box Monte Carlo (BB-MC) method \cite{MC-divergence} for estimating the divergence $\gamma_B^t$ from \eqref{eq:gamma_B_general}. To construct an estimate, the BB-MC method executes the denoiser again $\bg_B(\xAB^t + \epsilon \be)$ with the original input $\xAB^t$ perturbed by a random vector $\epsilon \be$, where each entry of $\be$ takes the values of $-1$ or $+1$ with equal probability, and we choose the scalar $\epsilon$ as in the GAMP library\footnote{The library is available on https://sourceforge.net/projects/gampmatlab/}

\noindent
\begin{equation*}
    \epsilon = 0.1 \min \big( \vAB^t, \inv{N} ||\xAB^t||_1 \big) + e
\end{equation*}

\noindent
where $e$ is the the float point precision in MATLAB. 

\begin{figure}
\centering
\includegraphics[width=0.25\textwidth]{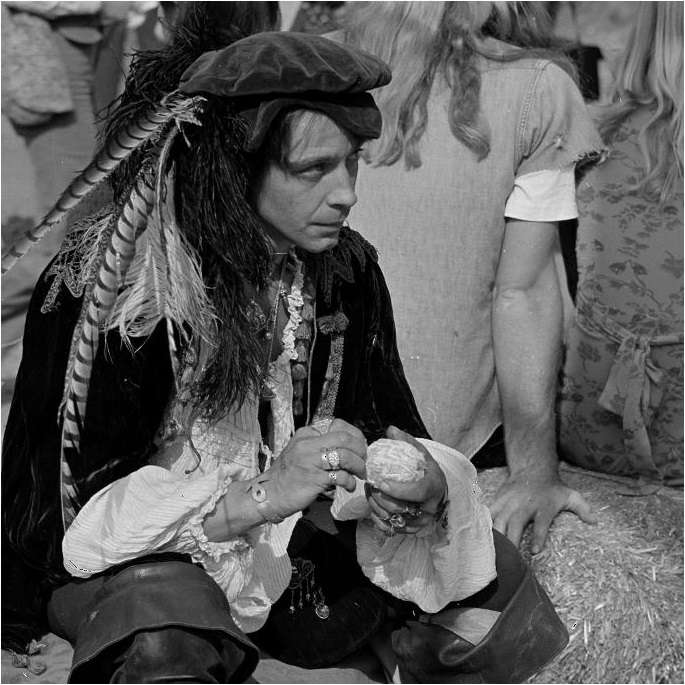}
\caption{The ground truth image to be reconstructed.}
\label{fig:ground_truth_image}
\end{figure}

For all the experiments, we will consider recovering a natural image 'man' of dimension $1024$ by $1024$ shown on Figure \ref{fig:ground_truth_image} and measured by a FIJL operator with the condition number $\kappa(\bA) = 1000$ (unless stated otherwise) and subsampling factor $\delta = \frac{M}{N} = 0.05$. Lastly, we set the measurement noise variance $v_w$ to achieve SNR $\frac{||\bx||^2}{||\bw||^2}$ of $40dB$. 

\subsection{Comparison of different CG-VAMP algorithms}

In our first experiment, we numerically study the stability of CG-VAMP with the correction method \eqref{eq:zero_init_CG_gamma_A} and the variance $\vAB^{t,i}$ update based on the SE \eqref{eq:zero_init_CG_v_AB} proposed in \cite{CG_EP}. We refer to this algorithm as \textit{CG-VAMP A}. We compare CG-VAMP A against CG-VAMP with the proposed correction method \eqref{eq:iterative_gamma_zero_init_CG} and the variance $\vAB^{t,i}$ estimator based on \eqref{eq:v_ab_efficient}, which will be referred as \textit{CG-VAMP B}. We compare the algorithms in the setting described in Section \ref{sec:numerical_experiments} and for all the algorithm we estimate $\vBA^t$ as

\noindent
\begin{equation}
    \hvBA^t = \inv{N} ||\bz_t||^2 - \delta \Hat{v}_w
\end{equation}

\noindent
which corresponds to \eqref{eq:psi_LSL_estimator} with $v_w$ replaced by $\hvw$. In the experiment, the CG-VAMP algorithms are not provided with the oracle information about the singular spectrum $\bS$ of $\bA$ and for both \eqref{eq:zero_init_CG_gamma_A} and \eqref{eq:zero_init_CG_v_AB} we estimate the moments $\chi_j$ by averaging over $1000$ Monte-Carlo trials proposed in the works \cite{MC_matrix_moments_1}, \cite{MC_matrix_moments_2}, \cite{MC-divergence}. To compare the performance, we compute the oracle Normalized Mean Squared Error (NMSE) $\frac{||\bg_B(\xAB^t) - \bx||^2}{||\bx||^2}$ of the algorithms.

\begin{figure}
\centering
\includegraphics[width=0.4\textwidth]{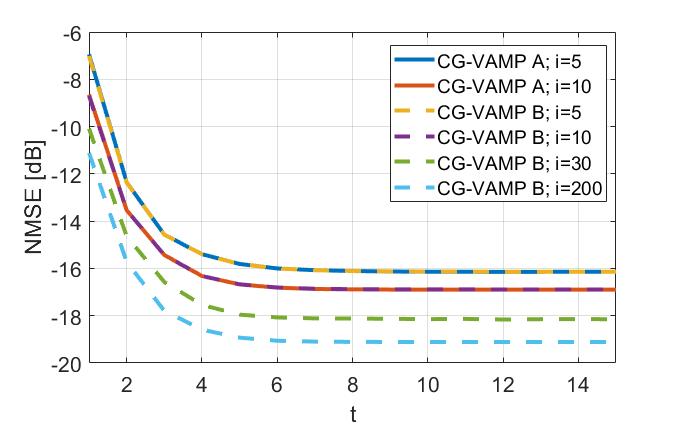}
\caption{NMSE versus outer-loop iteration number $t$. The dashed curves correspond to CG-VAMP A, while the solid lines represent the proposed CG-VAMP B.}
\label{fig:CG_VAMP_A_vs_CG_VAMP_B}
\end{figure}

The NMSE of CG-VAMP A and CG-VAMP B with different $i$ averaged over $10$ realizations is shown on Figure \ref{fig:CG_VAMP_A_vs_CG_VAMP_B}, where we do not plot those algorithms that diverge immediately, as in the case of CG-VAMP A that becomes unstable for $i=15$ and above. As seen from the plot, the practical version of CG-VAMP B remains stable for a wide range of inner-loop iterations and provides a progressively better estimate as $i$ increases.

\subsection{CG-VAMP with Adaptive CG} \label{sec:ACG-VAMP experiments}

\begin{figure}
\centering
\includegraphics[width=0.4\textwidth]{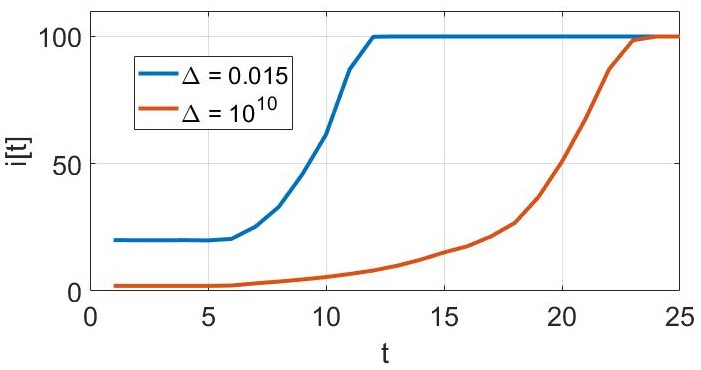}
\caption{The number of CG iterations $i[t]$ that follow only the stopping criterion \eqref{eq:CG_stopping_rule} (red curve) and that follows both \eqref{eq:CG_stopping_rule} and \eqref{eq:CG_stopping_rule_2} (blue curve).}
\label{fig:ACG_i_t}
\end{figure}

\begin{figure}
\centering
\includegraphics[width=0.5\textwidth]{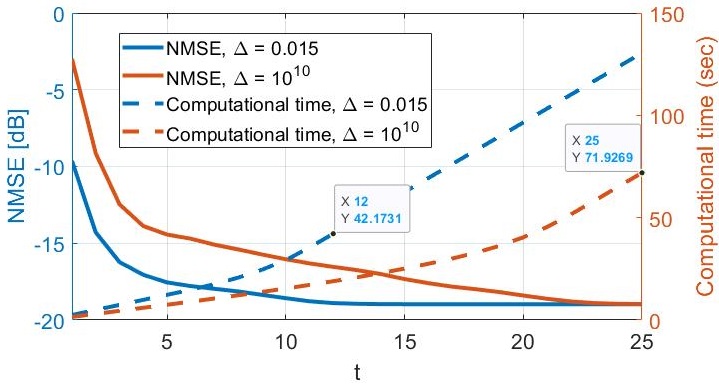}
\caption{The NMSE and the computational time of CG-VAMP with ACG with two types of stopping criteria.}
\label{fig:ACG_NMSE_time}
\end{figure}

In the next experiment we demonstrate the importance of choosing the right number of CG iterations $i$ at each outer-loop iteration $t$. In this experiment we use the CG-VAMP algorithm with the correction method \eqref{eq:iterative_gamma_zero_init_CG} and use ACG algorithm instead of the regular CG. To assess the effect of the stopping criteria \eqref{eq:CG_stopping_rule} and \eqref{eq:CG_stopping_rule_2}, we choose the expected variance $\vAB^{t,i}$ reduction constant $c = 0.9$, the threshold for the normalized reduction of the variance $\vAB^{t,i}$ per inner-loop iteration $\Delta = 0.015$ and the maximum number of ACG iterations $i_{max} = 100$. In this setting, we confirm the results from \cite{OurPaper} and demonstrate that when the algorithm progresses and reduces its intrinsic variance, the number of CG iterations required to preserve the same level of per-iteration improvement changes. To do this, we plotted $i[t]$ versus $t$ on Figure \ref{fig:ACG_i_t}, where the red curve represents ACG with only the stopping criterion \eqref{eq:CG_stopping_rule}, while \eqref{eq:CG_stopping_rule_2} is effectively disabled by setting $\Delta = 10^{10}$. The blue curve on the same plot represents $i[t]$ that follows both stopping criteria \eqref{eq:CG_stopping_rule} and \eqref{eq:CG_stopping_rule_2} with $\Delta = 0.015$ and leads to a faster time-wise convergence as shown on Figure \ref{fig:ACG_NMSE_time}. Here we would like to emphasize that in this experiment we chose a relatively fast\footnote{In this experiment we implemented the DnCNN with a GPU acceleration.} denoiser so the computational time improvement gained by adding the stopping criterion \eqref{eq:CG_stopping_rule_2} is only roughly $60\%$, while if we chose a more expensive denoiser like BM3D, the computational time difference of the two algorithm would be much higher.

\subsection{Practical implementations of WS-CG-VAMP}

Next, we numerically compare several practical versions of WS-CG-VAMP that utilize different types of correction scalars $\gamma_A^{t,\tau}$ that were discussed in Sections \ref{sec:CG_VAMP_LSL}-\ref{sec:stable_estimation_in_CG_VAMP} and \ref{subsec:warm-starting_appropriate_correction}-\ref{subsec:single_correction_WS_CG_VAMP}. To improve the stability of algorithms, all WS-CG-VAMP use the variance $\vAB^{t,i}$ estimator based on \eqref{eq:v_ab_efficient}. In this setup, we compare two pairs of WS-CG-VAMP. The first pair uses a single correction scalar $\hgam_A^{t,i}$ and the full set of correction scalars $\{\hgam_A^{t,\tau,i}\}_{\tau=0}^t$ estimated based on \eqref{eq:zero_init_CG_gamma_A} and \eqref{eq:WS_CG_correction_scalars} respectively. For these two algorithms, we estimate the moments $\chi_j$ as in the first experiment and we refer to these algorithms as \textit{WS-CG-VAMP A.single} and \textit{WS-CG-VAMP A.full}. In the same way, we have implemented a pair of WS-CG-VAMP with a single and the full set of correction scalars estimated based on \eqref{eq:iterative_gamma_zero_init_CG} and \eqref{eq:iterative_gamma_WS_CG} respectively. We refer to them as \textit{WS-CG-VAMP B.single} and \textit{WS-CG-VAMP B.full}, and for CG-VAMP B.full we estimate the correlation measures $\psi_{\tau, \tau'}$ for $\tau, \tau' \leq t$ through \eqref{eq:psi_LSL_estimator}. Additionally, we implement CG-VAMP B and VAMP for comparison. We evaluate the performance of all the algorithms for different condition numbers $\kappa(\bA) = \{10^2, 10^3, 10^4\}$.

\noindent
\begin{figure*}
\centering
\includegraphics[width=1\textwidth]{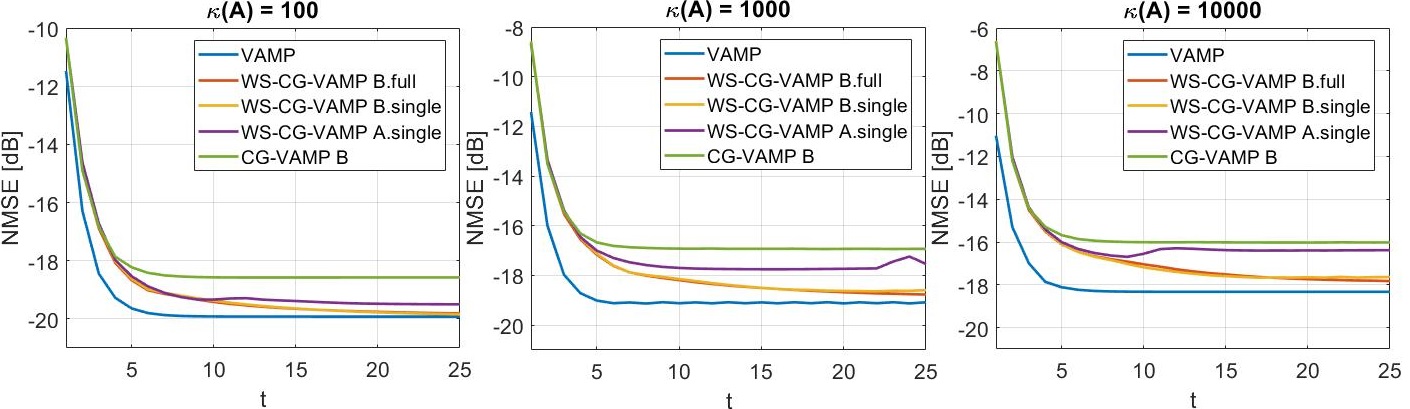}
\caption{The NMSE of VAMP, CG-VAMP B and different versions of WS-CG-VAMP versus outer-loop iteration $t$.}
\label{fig:practical_WS_CG_VAMP_NMSE}
\end{figure*}

In this context, WS-CG-VAMP B.full diverges at iteration $t=4$ for the three condition numbers, which we attribute to the fact that this estimator is sensitive to the error in estimating the moments $\chi_j$ and variances $v_w$ and $\{\vBA^{\tau}\}_{\tau = 0}^t$. The rest algorithms remain stable throughout iterations and their NMSE averaged over $10$ realizations is shown on Figure \ref{fig:practical_WS_CG_VAMP_NMSE}. As seen from the plot, WS-CG-VAMP B.single and WS-CG-VAMP B.full demonstrate almost identical performance and a consistent progression over $t = (1,...,25)$, although at some point we observe (not shown here) that WS-CG-VAMP B.single diverges, which we attribute to accumulation of the correlation between $\bh_t$ and $\bq_t$ as discussed in Section \ref{subsec:single_correction_WS_CG_VAMP}. The WS-CG-VAMP A.single is stable as well, although its fixed point is inferior to those of the previous two algorithms. The inferior reconstruction result might be partially explained by the fact that the estimator \eqref{eq:zero_init_CG_gamma_A} represents a black-box method that is blind to the actual data in the algorithm, contrary to \eqref{eq:iterative_gamma_zero_init_CG} that iteratively extracts the information from the vectors $\bz_t$ and $\bmu_A^{t,i}$ and might, potentially, account for the innacuracy of the SE. Lastly, we see from the plot that provided the same amount of resources, the WS-CG-VAMP algorithm gets much closer to the fixed point of VAMP compared to CG-VAMP.

\subsection{The SE for WS-CG in WS-CG-VAMP}

\begin{figure}
\centering
\includegraphics[width=0.5\textwidth]{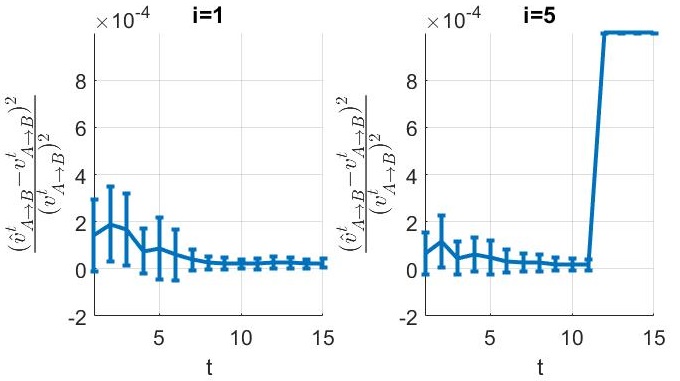}
\caption{The normalized error of the estimator based on \eqref{eq:v_ab_WS_CG} in WS-CG-VAMP with $i=1$ and $i=5$ respectively. The main line corresponds to the mean of the normalized error averaged over $10$ realizations, while the error bars corresponds to the standard deviation. The error is clipped to $10^{-4}$.}
\label{fig:v_ab_SE_error}
\end{figure}

Next, we would like to confirm the validity of the evolution model \eqref{eq:v_ab_WS_CG} of $\vAB^{t,i}$ in the WS-CG-VAMP algorithm. To measure the accuracy of the estimate $\hvAB^{t,i}$, we compute the oracle variance $\vAB^{t,i}$ and evaluate the corresponding normalized error $\frac{(\vAB^{t,i} - \hvAB^{t,i})^2}{(\vAB^{t,i})^2}$. We consider two cases: when $i=1$ and when $i=5$. For $i=1$, the normalized error of estimating $\vAB^{t,i}$ averaged over $10$ realizations is shown to the left on Figure \ref{fig:v_ab_SE_error}. As seen from the plot, the estimator of $\vAB^{t,i}$ based on \eqref{eq:v_ab_WS_CG} accurately predicts the true magnitude  of the error $\inv{N} ||\bh_t||^2$. For the next experiment where we set $i=5$, we increase the precision of MATLAB calculations to $64$ digits instead of the standard $16$, since the maximum order of the spectral moments grows as $t^2 i^2 + 2$ and those numbers quickly grow beyond the standard precision. For WS-CG-VAMP with $i=5$, the accuracy of the SE's prediction is shown to the right on Figure \ref{fig:v_ab_SE_error}, where at iteration $13$ the estimator diverges due to lack of precision (this can be solved by using more than $64$ digits).

\section{Conclusions}

Vector Approximate Message Passing (VAMP) \cite{VAMP}, \cite{NS-VAMP} demonstrates great reconstruction performance on linear inverse problems, but its applicability is limited to relatively small dimensional signals due to the necessity in computing the expensive LMMSE estimator. In this work we considered an extension of the VAMP algorithm to Conjugate Gradient VAMP (CG-VAMP) and derived asymptotically rigorous scalable methods for correcting, tuning and accelerating the CG algorithm within CG-VAMP. The CG-VAMP algorithm with the proposed Adaptive CG (ACG) algorithm demonstrates consistent and fast reconstruction of large-scale signals measured by highly ill-conditioned and underdetermined operators. 

In the second part of the work we considered the potential of warm-starting the CG algorithm within CG-VAMP. We derived the SE of  Warm-Started CG-VAMP (WS-CG-VAMP) and developed the asymptotic closed-form solution for the multidimensional correction of Warm-Started CG. Additionally, we proposed asymptotically rigorous and stable methods for implementing WS-CG-VAMP for solving large scale image reconstruction problems. The numerical experiments demonstrate that WS-CG-VAMP with a few CG iterations is able to get close to the VAMP fixed point and substantially outperform the regular CG-VAMP. Yet, we have observed that when WS-CG-VAMP is executed for large $t$, the error of the estimator \eqref{eq:psi_LSL_estimator} for $\Psi_t$ becomes larger and leads to poor performance of the overall WS-CG-VAMP algorithm. In future work we plan to investigate whether there are more robust ways for estimating the correlation matrix $\Psi_t$ within the family of Message-Passing algorithms. Additionally, we plan to consider how to extend the developed methods to the case where the covariance matrices of the cavity densities $\qAB^t$ and $\qBA^t$ from the section \ref{sec:EP_background} are block-isotropic as considered in the recent paper on \textit{Variable Density Approximate Message Passing} \cite{VDAMP}.

\section{Acknowledgements}

The authors would like to thank Arian Maleki and the anonymous reviewers for the valuable suggestions that have helped to improve the quality of the work significantly.

\section{Appendix A}
\begin{proof}[\unskip\nopunct]

In the following, we derive the mapping \eqref{eq:bz_tau_to_bmu} for the WS-CG algorithm

\noindent
\begin{gather}
    \bmu_A^{t,i} = \bmu_A^{t,i-1} + a_t^{i-1} \bp_t^{i-1} \label{eq:bmu}\\
    \bp_t^i = \bz_t - \bW_t \bmu_A^{t,i} + b_t^{i-1} \bp_t^{i-1} \label{eq:bp}
\end{gather}

\noindent
where $a_t^i$ and $b_t^i$ are computed as in Algorithm 2 and we have the initializations \eqref{eq:mu_init}-\eqref{eq:p_init} for $t \geq 1$. For $t=0$, when we use zero initializations for CG, the mapping \eqref{eq:bz_tau_to_bmu} was proved in \cite{CG_EP}. For $t \geq 1$, the following analysis is based on the observation that given the sets of scalars $\big\{ a_t^j \big\}_{j=0}^i$ and $\big\{ b_t^j \big\}_{j=0}^i$, the updates \eqref{eq:bmu}-\eqref{eq:bp} become linear mappings. Because they are linear, it is possible to separate the impact of each input vector $\bz_t$, $\bmu_A^{t-1,i}$ and $\bp_{t-1}^{i-1}$ on $\bmu_A^{t,i}$ and $\bp_t^i$ and reformulate \eqref{eq:bmu}-\eqref{eq:bp} as

\noindent
\begin{gather}
    \bmu_A^{t,i} = \bF_{\bz}^{t,i}(\bz_t) + \bF_{\bmu}^{t,i}(\bmu_A^{t-1,i}) + \bF_{\bp}^{t,i}(\bp_{t-1}^{i-1}) \label{eq:bmu_split} \\
    \bp_t^i = \bG_{\bz}^{t,i}(\bz_t) + \bG_{\bmu}^{t,i}(\bmu_A^{t-1,i}) + \bG_{\bp}^{t,i}(\bp_{t-1}^{i-1}) \label{eq:bp_split}
\end{gather}

\noindent
where $\bF_{(\cdot)}^{t,i}(\cdot)$ and $\bG_{(\cdot)}^{t,i}(\cdot)$ are some functions and we omitted their dependence on $\big\{ a_t^j \big\}_{j=0}^i$ and $\big\{ b_t^j \big\}_{j=0}^i$. The following lemma defines the recursive structure of these functions.

\begin{lemma} \label{lemma:bmu_bp_recursive_structure}

    For the WS-CG algorithm \eqref{eq:bmu}-\eqref{eq:bp} with the initializations \eqref{eq:mu_init}-\eqref{eq:p_init}, we have:
    
    \begin{itemize}[leftmargin=3mm]
        
        \item For $\bF_{\bz}^{t,i}$ and $\bG_{\bz}^{t,i}$
        
        \noindent
        \begin{gather}
            \bF_{\bz}^{t,i}(\bz_t) = \bF_{\bz}^{t,i-1}(\bz_t) + a_t^{i-1} \bG_{\bz}^{t,i-1}(\bz_t) \label{eq:bmu_from_bz}\\
            \bG_{\bz}^{t,i}(\bz_t) = \bz_t - \bW_t \bF_{\bz}^{t,i}(\bz_t) + b_t^{i-1} \bG_{\bz}^{t,i-1}(\bz_t) \label{eq:bp_from_bz}
        \end{gather}
        
        \noindent
        with $\bF_{\bz}^{t,0}(\bz_t) = \bzero_M$ and $\bG_{\bz}^{t,0}(\bz_t) = \bz_t$.

        \item For $\bF_{\bmu}^{t,i}$ and $\bG_{\bmu}^{t,i}$
        
        \noindent
        \begin{gather}
            \bF_{\bmu}^{t,i}(\bmu_A^{t-1,i}) = \bF_{\bmu}^{t,i-1}(\bmu_A^{t-1,i}) + a_t^{i-1} \bG_{\bmu}^{t,i-1}(\bmu_A^{t-1,i}) \label{eq:bmu_from_bmu}\\
            \bG_{\bmu}^{t,i}(\bmu_A^{t-1,i}) = - \bW_t \bF_{\bmu}^{t,i}(\bmu_A^{t-1,i}) + b_t^{i-1} \bG_{\bmu}^{t,i-1}(\bmu_A^{t-1,i}) \label{eq:bp_from_bmu}
        \end{gather}
        
        \noindent
        with $\bF_{\bmu}^{t,0}(\bmu_A^{t-1,i}) = \bmu_A^{t-1,i}$ and $\bG_{\bmu}^{t,0}(\bmu_A^{t-1,i}) = - \bW_t \bmu_A^{t-1,i}$.

        \item For $\bF_{\bp}^{t,i}$ and $\bG_{\bp}^{t,i}$
        
        \noindent
        \begin{gather}
            \bF_{\bp}^{t,i}(\bp_{t-1}^{i-1}) = \bF_{\bp}^{t,i-1}(\bp_{t-1}^{i-1}) + a_t^{i-1} \bG_{\bp}^{t,i-1}(\bp_{t-1}^{i-1}) \label{eq:bmu_form_bp}\\
            \bG_{\bp}^{t,i}(\bp_{t-1}^{i-1}) = - \bW_t \bF_{\bp}^{t,i}(\bp_{t-1}^{i-1}) + b_t^{i-1} \bG_{\bp}^{t,i-1}(\bp_{t-1}^{i-1}) \label{eq:bp_from_bp}
        \end{gather}
        
        \noindent
        with $\bF_{\bp}^{t,0}(\bp_{t-1}^{i-1}) = \bzero_M$ and $\bG_{\bp}^{t,0}(\bp_{t-1}^{i-1}) = b_{t-1}^{i-1} \bp_{t-1}^{i-1}$
        
    \end{itemize}

\end{lemma}

\begin{proof}
    We prove this lemma only for the mappings $\bF_{\bz}^{t,i}(\bz_t)$ and $\bG_{\bz}^{t,i}(\bz_t)$, since the proof for the rest is similar. The proof is by induction and begins with $i=0$, for which we have 
    
    \noindent
    \begin{gather}
        \bmu_A^{t,0} = 0 \cdot \bz_t + \bmu_A^{t-1,i} + 0 \cdot \bp_{t-1}^{i-1} \\
        \bp_t^0 = \bz_t - \bW_t \bmu_A^{t-1,i} + b_{t-1}^i \bp_{t-1}^{i-1}
    \end{gather}
    
    \noindent
    as follows from \eqref{eq:mu_init} and \eqref{eq:p_init}. Therefore, if we define
    
    \noindent
    \begin{equation}
        \bF_{\bz}^{t,0}(\bz_t) = \bzero_M \quad \quad \bG_{\bz}^{t,0}(\bz_t) = \bz_t
    \end{equation}
    
    \noindent
    then the functions $\bF_{\bz}^{t,0}$ and $\bG_{\bz}^{t,0}$ fully cover the dependence of $\bmu_A^{t,0}$ and $\bp_t^0$ from \eqref{eq:bmu_split} and \eqref{eq:bp_split} on $\bz_t$. Next we assume \eqref{eq:bmu_from_bz} and \eqref{eq:bp_from_bz} hold up to $i = k$ and we prove it for $i = k+1$. Since \eqref{eq:bmu} is linear with respect to $\bmu_A^{t,i-1}$ and $\bp_t^{i-1}$ given the scalar $a_t^k$, we have that
    
    \begin{equation}
        \bF_{\bz}^{t,k+1}(\bz_t) = \bF_{\bz}^{t,k}(\bz_t) + a_t^{k} \bG_{\bz}^{t,k}(\bz_t)
    \end{equation}
    
    \noindent
    which confirms \eqref{eq:bmu_from_bz}. In the same way we prove \eqref{eq:bp_from_bz}. 
    \end{proof}

The above lemma implies that we can analyze the impact of each input vector $\bz_{t-1}$, $\bmu_A^{t-1,i}$ and $\bp_{t-1}^{i-1}$ on the output vectors $\bmu_A^{t,i}$ and the conjugate direction vector $\bp_t^i$ separately. Additionally, one might notice that given the sets $\big\{ a_t^j \big\}_{j=0}^i$ and $\big\{ b_t^j \big\}_{j=0}^i$, the iterations  \eqref{eq:bmu_split}-\eqref{eq:bp_split} represent linear mappings. To establish the exact structure of these mappings, we define three pairs of scalar sequences. The first one is

\noindent
\begin{gather}
    f_{\bz}^{t,i}[j] = f_{\bz}^{t,i-1}[j] + a_t^i g_{\bz}^{t,i-1}[j] \label{eq:f_t_z} \\
    g_{\bz}^{t,i}[j] = \delta_{j,0} - v_w f_{\bz}^{t,i}[j] - \vBA^t f_{\bz}^{t,i}[j-1] + b_t^i g_{\bz}^{t,i-1}[j] \label{eq:g_t_z}
\end{gather}

\noindent
with $f_{\bz}^{t,0}[0] = 0$, $g_{\bz}^{t,0}[0] = 1$, $f_{\bz}^{t,i}[j] = 0$ and $g_{\bz}^{t,i}[j] = 0$ for $j \not\in \{0,...,i\}$ and with $\delta_{j,0}$ defining the Kronecker delta. The second sequence corresponds to

\noindent
\begin{gather}
    f_{\bmu}^{t,i}[j] = f_{\bmu}^{t,i-1}[j] + a_t^i g_{\bmu}^{t,i-1}[j] \label{eq:f_t_mu} \\
    g_{\bmu}^{t,i}[j] = -v_w f_{\bmu}^{t,i}[j] - \vBA^t f_{\bmu}^{t,i}[j-1] + b_t^i g_{\bmu}^{t,i-1}[j] \label{eq:g_t_mu}
\end{gather}

\noindent
with $f_{\bmu}^{t,0}[0] = 1$, $g_{\bmu}^{t,0}[0] = -v_w$, $g_{\bmu}^{t,0}[1] = -\vBA^t$, $f_{\bmu}^{t,i}[j] = 0$ for $j \not\in \{0,...,i\}$ and $g_{\bmu}^{t,i}[j] = 0$ for $j \not\in \{0,...,i+1\}$. Lastly, the third sequence is

\noindent
\begin{gather}
    f_{\bp}^{t,i}[j] = f_{\bp}^{t,i-1}[j] + a_t^i g_{\bp}^{t,i-1}[j] \label{eq:f_t_p} \\
    g_{\bp}^{t,i}[j] = -v_w f_{\bp}^{t,i}[j] - \vBA^t f_{\bp}^{t,i}[j-1] + b_t^i g_{\bp}^{t,i-1}[j] \label{eq:g_t_p}
\end{gather}

\noindent
with $f_{\bp}^{t,0}[0] = 0$, $g_{\bp}^{t,0}[0] = b_{t-1}^{i-1}$, $f_{\bp}^{t,i}[j] = 0$ and $g_{\bp}^{t,i}[j] = 0$ for $j \not\in \{0,...,i\}$.

With these definitions, first, we present the alternative linear forms of $\bF_{\bz}^i(\bz_t)$ and $\bG_{\bz}^i(\bz_t)$ proved in \cite{CG_EP}.

\begin{lemma} \label{lemma:bF_and_bG_z}
    
    \textit{Lemma 1 in \cite{CG_EP}:} Let $\bPhi = \bA \bA^T$. Then $\bF_{\bz}^{t,i}(\bz_t)$ and $\bG_{\bz}^{t,i}(\bz_t)$ from \eqref{eq:bmu_from_bz} and \eqref{eq:bp_from_bz} can be alternatively represented as

    \noindent
    \begin{gather}
        \bF_{\bz}^{t,i}(\bz_{t}) = \sum_{j=0}^i f_{\bz}^{t,i}[j] \bPhi^j \bz_{t} \label{eq:F_z_linear}\\ 
        \bG_{\bz}^{t,i}(\bz_{t}) = \sum_{j=0}^i g_{\bz}^{t,i}[j] \bPhi^j \bz_{t} \label{eq:G_z_linear}
    \end{gather}

\end{lemma}

We follow the same idea and present the alternative mappings for the pairs $\bF_{\bmu}^i(\bmu_A^{t,i})$ and $\bG_{\bmu}^i(\bmu_A^{t,i})$, and $\bF_{\bp}^i(\bp_t^i)$ and $\bG_{\bp}^i(\bp_t^i)$.

\begin{lemma} \label{lemma:bF_and_bG_mu_p}
    
    Let $\bPhi = \bA \bA^T$. Then we have
    
    \noindent
    \begin{gather}
        \bF_{\bmu}^{t,i}(\bmu_A^{t-1,i}) = \sum_{j=0}^i f_{\bmu}^{t,i}[j] \bPhi^j  \bmu_A^{t-1,i} \label{eq:F_mu_linear}\\
        \bG_{\bmu}^{t,i}(\bmu_A^{t-1,i}) = \sum_{j=0}^{i+1} g_{\bmu}^{t,i}[j] \bPhi^j \bmu_A^{t-1,i} \label{eq:G_mu_linear} \\
        \bF_{\bp}^{t,i}(\bp_{t-1}^{i-1}) = \sum_{j=0}^i f_{\bp}^{t,i}[j] \bPhi^j  \bp_{t-1}^{i-1} \label{eq:F_p_linear}\\
        \bG_{\bp}^{t,i}(\bp_{t-1}^{i-1}) =  \sum_{j=0}^i g_{\bp}^{t,i}[j] \bPhi^j \bp_{t-1}^{i-1} \label{eq:G_p_linear}
    \end{gather}
    
\end{lemma}

\begin{proof}
    See Appendix B.
\end{proof}

Lemmas \ref{lemma:bF_and_bG_z} and \ref{lemma:bF_and_bG_mu_p} together with \eqref{eq:bmu_split}-\eqref{eq:bp_split} imply that the CG output $\bmu_A^{t,i}$ and the vector $\bp_t^i$ can be represented as linear mappings of the vectors $\bz_t$, $\bmu_A^{t-1,i}$ and $\bp_{t-1}^{i-1}$ given the set of scalars $\{a_t^j\}_{j=0}^i$ and $\{b_t^j\}_{j=0}^i$ as

\noindent
\begin{gather}
    \bmu_A^{t,i} = \bF_{\bz}^{t,i} \bz_t + \bF_{\bmu}^{t,i} \bmu_A^{t-1,i} + \bF_{\bp}^{t,i} \bp_{t-1}^{i-1} \label{eq:btmu_linear} \\
    \bp_t^i = \bG_{\bz}^{t,i} \bz_t + \bG_{\bmu}^{t,i} \bmu_A^{t-1,i} + \bG_{\bp}^{t,i} \bp_{t-1}^{i-1} \label{eq:btp_linear}
\end{gather}
\noindent
Now we can finish the derivation of Theorem \ref{theorem:bz_tau_to_mu} in two steps. First, we repeatedly apply Lemma \ref{lemma:bF_and_bG_mu_p} to \eqref{eq:btmu_linear} to obtain a mapping from the set $\big\{ \bz_{\tau} \big\}_{\tau=0}^t$ to $\bmu_A^{t,i}$. And next, we group up the terms that scale $\bPhi$ with the same power $k$. For this, we present two sequences $r_{\tau}^{t,d}[k]$ and $u_{\tau}^{t,d}[k]$ that for $0 \leq \tau < t$ are defined as
    
\noindent
\begin{equation}
    r_{\tau}^{t,d}[k] = \sum_{j=0}^{d}  f_{\bmu}^{t,d}[j] r_{\tau}^{t-1,d}[k-j] + f_{\bp}^{t,d}[j] u_{\tau}^{t-1,i-1}[k-j] \label{eq:r_t_lemma}
\end{equation}

\noindent
\begin{equation}
    u_{\tau}^{t,d}[k] = \sum_{j=0}^{d} g_{\bmu}^{t,d}[j] r_{\tau}^{t-1,d}[k-j] + g_{\bp}^{t,d}[j] u_{\tau}^{t-1,i-1}[k-j]  \label{eq:u_t_lemma}
\end{equation}

\noindent
with $r_{\tau}^{t,i}[k]=0$ and $u_{\tau}^{t,i}[k]=0$ for $k \not\in \{0,...,(t-\tau)i\}$. For $\tau=t$ we set $r_t^{t,i}[k] = f_{\bz}^{t,i}[k]$ and $u_t^{t,i}[k] = g_{\bz}^{t,i}[k]$. With these definitions, we can show by induction that 

\noindent
\begin{gather}
    \bmu_A^{t,i} = \sum_{\tau = 0}^{t} \sum_{k = 0}^{(t - \tau) i} r_{\tau}^{t,i}[k] \bPhi^k \bz_{\tau} \label{eq:bz_tau_to_bmu_proof}\\
    \bp_t^i = \sum_{\tau = 0}^{t} \sum_{k = 0}^{(t - \tau) i} u_{\tau}^{t,i}[k] \bPhi^k \bz_{\tau} \label{eq:bz_tau_to_bp_proof}
\end{gather}

\noindent
where \eqref{eq:bz_tau_to_bmu_proof} is related to \eqref{eq:bz_tau_to_bmu} through the SVD of $\bPhi = \bA \bA^T = \bU \bLambda \bU^T$. We start the induction with $t=0$, for which WS-CG and the regular CG are the same, and in \eqref{eq:btmu_linear}-\eqref{eq:btp_linear} we only have the mappings $\bF_{\bz}^{t,i}$ and $\bG_{\bz}^{t,i}$. In this case, by setting $r_0^{0,i}[k] = f_{\bz}^{0,i}[k]$ and $u_0^{0,i}[k] = g_{\bz}^{0,i}[k]$, and following Lemma \ref{lemma:bF_and_bG_z}, we confirm \eqref{eq:bz_tau_to_bmu_proof} and \eqref{eq:bz_tau_to_bp_proof} for $t=0$.

Next, assume \eqref{eq:bz_tau_to_bmu_proof} and \eqref{eq:bz_tau_to_bp_proof} hold up to iteration $t$ and we would like to confirm them for the iterations $t+1$. The mapping from $\bz_{t+1}$ to $\bmu_A^{t+1,i}$ can be confirmed by setting $r_{t+1}^{t+1,i}[k] = f_{\bz}^{t+1,i}[k]$ and applying Lemma \ref{lemma:bF_and_bG_z} to \eqref{eq:btmu_linear}. For the mapping from $\big\{ \bz_{\tau} \big\}_{\tau=0}^t$ to $\bmu_A^{t+1,i}$ we apply the results from Lemma \ref{lemma:bF_and_bG_mu_p} to \eqref{eq:bz_tau_to_bmu_proof} and \eqref{eq:bz_tau_to_bp_proof} to obtain

\noindent
\begin{align}
    &\bF_{\bmu}^{t+1,i} \bmu_{t}^i + \bF_{\bp}^{t+1,i} \bp_{t}^{i-1} = \bF_{\bmu}^{t+1,i} \sum_{\tau = 0}^{t} \sum_{k = 0}^{(t - \tau) i} r_{\tau}^{t,i}[k] \bPhi^k \bz_{\tau} \nonumber\\
    &+ \bF_{\bp}^{t+1,i} \sum_{\tau = 0}^{t} \sum_{k = 0}^{(t - \tau) i} u_{\tau}^{t,i-1}[k] \bPhi^k \bz_{\tau} \nonumber\\
    &= \sum_{\tau = 0}^{t} \sum_{k = 0}^{(t - \tau) i} \sum_{j=0}^i f_{\bmu}^{t,i}[j] \bPhi^j r_{\tau}^{t,i}[k] \bPhi^k \bz_{\tau} \nonumber\\
    &+ \sum_{\tau = 0}^{t} \sum_{k = 0}^{(t - \tau) i} \sum_{j=0}^{i} f_{\bp}^{t,i}[j] \bPhi^j u_{\tau}^{t,i-1}[k] \bPhi^k \bz_{\tau} \nonumber\\
    &= \sum_{\tau = 0}^{t} \sum_{k = 0}^{(t - \tau) i} \sum_{j=0}^{i} \Big( f_{\bmu}^{t,i}[j] r_{\tau}^{t,i}[k] + f_{\bp}^{t,i}[j] u_{\tau}^{t,i-1}[k] \Big) \bPhi^{k+j} \bz_{\tau} \nonumber\\
    &= \sum_{\tau = 0}^{t} \sum_{k = 0}^{(t - \tau) i} \sum_{j=0}^{i} \bphi_{\tau}^{t,i}[j,k] \bPhi^{k+j} \bz_{\tau} \label{eq:z_tau_to_F_mu_plus_F_p}
\end{align}

\noindent
where we defined an auxiliary scalar function

\noindent
\begin{equation}
    \phi_{\tau}^{t,i}[j,k] = f_{\bmu}^{t,i}[j] r_{\tau}^{t,i}[k] + f_{\bp}^{t,i}[j] u_{\tau}^{t,i-1}[k] \label{eq:bphi}
\end{equation}

\noindent
Note that the functions $f_{(\cdot)}^{t,i}[j]$ and $g_{(\cdot)}^{t,i}[j]$ are finite for finite $i$ since $a_t^i$, $b_t^i$, $v_w$ and $\vBA^t$ are finite and \eqref{eq:f_t_z}-\eqref{eq:g_t_p} involve only summation and subtraction operations. Then, because $r_{\tau}^{t,i}[k]$ and $u_{\tau}^{t,i}[k]$ are linear combinations of $f_{(\cdot)}^{t,i}[j]$ and $g_{(\cdot)}^{t,i}[j]$, we have that $\phi_{\tau}^{t,i}[j,k]$ is also finite. Therefore, we can change the order of summation with respect to $k$ and $j$ in \eqref{eq:z_tau_to_F_mu_plus_F_p} given $i$ is finite. Then, by the change of variables $k' = k + j$, we can arrive at 

\noindent
\begin{equation}
    \bF_{\bmu}^{t+1,i} \bmu_{t}^i + \bF_{\bp}^{t+1,i} \bp_{t}^{i-1} = \sum_{\tau = 0}^{t} \sum_{j=0}^{i} \sum_{k' = j}^{(t - \tau) i + j} \phi_{\tau}^{t,i}[j,k'-j] \bPhi^{k'} \bz_{\tau} \label{eq:z_tau_to_mu_pre_last_step}
\end{equation}

\noindent
Next, note that from the definition of $r_{\tau}^{t,i}[k]$ and $u_{\tau}^{t,i}[k]$ we have that $\phi_{\tau}^{t,i}[j,k]=0$ for $k \not\in \{0,...,(t-\tau)i\}$, which implies that in the last result we can set the lower bound of the summation for $k'$ to zero without affecting the result. Similarly, we can increase the upper bound of the same summation. Because in \eqref{eq:z_tau_to_F_mu_plus_F_p} the maximum value of $j$ is $i$, we change the upper bound of the summation for $k'$ to $(t - \tau) i + i = (t+1 - \tau) i$. With these changes we obtain 

\noindent
\begin{equation*}
    \bF_{\bmu}^{t+1,i} \bmu_{t}^i + \bF_{\bp}^{t+1,i} \bp_{t}^{i-1} = \sum_{\tau = 0}^{t} \sum_{j=0}^{i} \sum_{k' = 0}^{(t + 1 - \tau) i} \phi_{\tau}^{t,i}[j,k'-j] \bPhi^{k'} \bz_{\tau}
\end{equation*}

\noindent
Lastly, by changing the order of the summation with respect to $j$ and $k'$ and defining

\noindent
\begin{equation}
    r_{\tau}^{t,i}[k'] = \sum_{j=0}^{i} \phi_{\tau}^{t,i}[j,k'-j]
\end{equation}

\noindent
we arrive at \eqref{eq:bz_tau_to_bmu_proof} for $t+1$.

Because the structure of $r_{\tau}^{t,i}[k]$ and of $u_{\tau}^{t,i}[k]$ are similar and because the structure of the updates for $\bmu_A^{t,i}$ and for $\bp_t^i$ are the same, the proof of \eqref{eq:bz_tau_to_bp_proof} follows exactly the same steps as for proving \eqref{eq:bz_tau_to_bmu_proof} considered above. \end{proof}

\section*{Appendix B}
\begin{proof}[\unskip\nopunct]

The proof of Lemma \ref{lemma:bF_and_bG_mu_p} is done by induction. First, we consider the recursion \eqref{eq:bmu_from_bmu}-\eqref{eq:bp_from_bmu} and prove the mappings \eqref{eq:F_mu_linear}-\eqref{eq:G_mu_linear}. We begin the induction with $i=0$, for which we set $f_{\bmu}^{t,0}[0] = 1$ and use the right hand side of \eqref{eq:F_mu_linear} to obtain

\noindent
\begin{equation}
    f_{\bmu}^{t,0}[0] \bPhi^{0} \bmu_A^{t-1,i} = \bmu_A^{t-1,i} \\
\end{equation}

\noindent
which matches the initialization $\bF_{\bmu}^{t,0}(\bmu_A^{t-1,i}) = \bmu_A^{t-1,i}$ from Lemma \ref{lemma:bmu_bp_recursive_structure}. Similarly, setting

\noindent
\begin{equation*}
    g_{\bmu}^{t,0}[0] = -v_w \quad \quad g_{\bmu}^{t,0}[1] = -\vBA^t
\end{equation*}

\noindent
and using the right hand side of \eqref{eq:G_mu_linear} gives

\noindent
\begin{align*}
    &g_{\bmu}^{t,0}[0] \bPhi^{0} \bmu_A^{t-1,i} + g_{\bmu}^{t,0}[1] \bPhi^{1} \bmu_A^{t-1,i} \nonumber\\
    &= - v_w g_{\bmu}^{t,0}[0] \bPhi^{0} \bmu_A^{t-1,i} - \vBA^t \bPhi^{1} \bmu_A^{t-1,i} = - \bW_t \bmu_A^{t-1,i} 
\end{align*}

\noindent
which matches the initialization $\bG_{\bmu}^{t,0}(\bmu_A^{t-1,i}) = - \bW_t \bmu_A^{t-1,i}$ from Lemma \ref{lemma:bmu_bp_recursive_structure}.

Next, we assume \eqref{eq:F_mu_linear}-\eqref{eq:G_mu_linear} hold up to $i=k$ and we aim to confirm it for $i=k+1$. From \eqref{eq:F_mu_linear} we have 

\noindent
\begin{align}
    &\sum_{j=0}^{k+1} f_{\bmu}^{t,k+1}[j] \bPhi^j  \bmu_A^{t-1,i} \nonumber\\
    &\overset{(a)}{=} \sum_{j=0}^{k+1} (f_{\bmu}^{t,k}[j] + a_t^k g_{\bmu}^{t,k}[j]) \bPhi^j \bmu_A^{t-1,i} \nonumber\\
    &\overset{(b)}{=} \sum_{j=0}^{k} f_{\bmu}^{t,k}[j]\bPhi^j \bmu_A^{t-1,i} + a_t^k \sum_{j=0}^{k+1} g_{\bmu}^{t,k}[j]) \bPhi^j \bmu_A^{t-1,i} \nonumber\\
    &\overset{(c)}{=} \bF_{\bmu}^{t,i}(\bmu_A^{t-1,i}) + a_t^k \bG_{\bmu}^{t,i}(\bmu_A^{t-1,i})
\end{align}

\noindent
where (a) comes from \eqref{eq:f_t_mu}, (b) is due to $f_{\bmu}^{t,k}[k+1] = 0$ by definition and (c) is by the induction hypothesis. By comparing this result to \eqref{eq:bmu_from_bmu}, we confirm the validity of \eqref{eq:F_mu_linear}. Similarly, from \eqref{eq:G_mu_linear} we have

\noindent
\begin{align}
    \sum_{j=0}^{k+2} &g_{\bmu}^{t,k+1}[j]  \bPhi^j \bmu_A^{t-1,i} = \sum_{j=0}^{k+2} \bigg(-v_w f_{\bmu}^{t,k+1}[j] \nonumber\\
    &- \vBA^t f_{\bmu}^{t,k+1}[j-1] + b_t^k g_{\bmu}^{t,k}[j]\bigg) \bPhi^j \bmu_A^{t-1,i} 
\label{eq:mu_to_btp_intermediate}
\end{align}

\noindent
Here we have that

\noindent
\begin{align}
    \sum_{j=0}^{k+2} f_{\bmu}^{t,i+1}[j]\bPhi^j \bmu_A^{t-1,i} &\overset{(a)}{=} \sum_{j=0}^{k+1} f_{\bmu}^{t,i+1}[j]\bPhi^j \bmu_A^{t-1,i} \nonumber\\
    &= \bF_{\bmu}^{t,i+1}(\bmu_A^{t-1,i}) \label{eq:mu_to_btp_first}
\end{align}

\noindent
where (a) is due to $f_{\bmu}^{t,i+1}[i+2] = 0$ as follows from the definition. Also

\noindent
\begin{align}
    \sum_{j=0}^{k+2} g_{\bmu}^{t,i}[j]\big) \bPhi^j \bmu_A^{t-1,i} &\overset{(a)}{=} \sum_{j=0}^{k+1} g_{\bmu}^{t,i}[j]\big) \bPhi^j \bmu_A^{t-1,i} \nonumber\\
    &= \bG_{\bmu}^{t,i}(\bmu_A^{t-1,i}) \label{eq:mu_to_btp_second}
\end{align}

\noindent
where (a) is because $g_{\bmu}^{t,k}[k+2]\big) = 0$ by definition. Lastly, by making the substitution $e = j-1$, we can obtain

\noindent
\begin{align}
    &\sum_{j=0}^{k+2} f_{\bmu}^{t,k+1}[j-1]\bPhi^j \bmu_A^{t-1,i} = \sum_{e=-1}^{k+1} f_{\bmu}^{t,k+1}[e]\bPhi^{e+1} \bmu_A^{t-1,i} \nonumber\\
    &\overset{(a)}{=} \bPhi \sum_{e=0}^{k+1} f_{\bmu}^{t,k+1}[e]\bPhi^{e} \bmu_A^{t-1,i} = \bPhi \bF_{\bmu}^{t,i+1}(\bmu_A^{t-1,i}) \label{eq:mu_to_btp_third}
\end{align}

\noindent
where (a) is due to $f_{\bmu}^{t,k+1}[-1] = 0$ by definition. Inserting \eqref{eq:mu_to_btp_first}, \eqref{eq:mu_to_btp_second} and \eqref{eq:mu_to_btp_third} into \eqref{eq:mu_to_btp_intermediate} gives

\noindent
\begin{align}
    &\sum_{j=0}^{k+2} g_{\bmu}^{t,k+1}[j] \bPhi^j \bmu_A^{t-1,i} = -v_w \bF_{\bmu}^{t,i+1}(\bmu_A^{t-1,i}) \nonumber\\
    &- \vBA^t \bPhi \bF_{\bmu}^{t,i+1}(\bmu_A^{t-1,i}) + b_t^k \bG_{\bmu}^{t,i}(\bmu_A^{t-1,i}) \nonumber\\
    &= -\bW_t \bF_{\bmu}^{t,i+1}(\bmu_A^{t-1,i}) + b_t^k \bG_{\bmu}^{t,i}(\bmu_A^{t-1,i}) \nonumber\\
    &= \bG_{\bmu}^{t,i+1}(\bmu_A^{t-1,i})
\end{align}

\noindent
which completes the proof for \eqref{eq:F_mu_linear}-\eqref{eq:G_mu_linear}. To prove a similar result for \eqref{eq:F_p_linear}-\eqref{eq:G_p_linear}, first, we note that the sequences \eqref{eq:f_t_p}-\eqref{eq:g_t_p} are exactly the same as \eqref{eq:f_t_mu}-\eqref{eq:g_t_mu} so the induction part of the proof will be exactly the same as for proving \eqref{eq:F_mu_linear}-\eqref{eq:G_mu_linear}. The only difference between the two sequences are the initializations. Following the same idea, we set $f_{\bp}^{t,0}[0] = 0$ and use the right hand side of \eqref{eq:F_p_linear} to show

\noindent
\begin{equation}
    f_{\bp}^{t,0}[0] \bPhi^0 \bp_{t-1}^{i-1} = \bzero_M
\end{equation}

\noindent
which matches the initialization $\bF_{\bp}^{t,0}(\bp_{t-1}^{i-1}) = \bzero_M$ from Lemma \ref{lemma:bmu_bp_recursive_structure}. Similarly, let $g_{\bp}^{t,0}[0] = b_{t-1}^{i-1}$ so that the right hand side of \eqref{eq:G_p_linear} is equal to

\noindent
\begin{align*}
    &g_{\bp}^{t,0}[0] \bPhi^{0} \bp_t^0 = b_{t-1}^{i-1} \bPhi^{0} \bp_{t-1}^{i-1} = b_{t-1}^{i-1} \bp_{t-1}^{i-1}
\end{align*}

\noindent
which is consistent with the initialization $\bG_{\bp}^{t,0}(\bp_{t-1}^{i-1}) = b_{t-1}^{i-1} \bp_{t-1}^{i-1}$ from Lemma \ref{lemma:bmu_bp_recursive_structure}. \end{proof}

\section*{Appendix C}
\begin{proof}[\unskip\nopunct]

Next we prove Theorem \ref{theorem:WS_CG_correction_and_SE}. We begin with deriving the asymptotic result for the correction scalar $\gamma_A^{t,\tau,i}$. By applying \eqref{eq:bz_tau_to_bmu} to \eqref{eq:WS_gamma_A_assymptotic} and noting that $\bz_{\tau} = \bw - \bA \bq_{\tau}$ we obtain

\noindent
\begin{align*}
    &\gamma_A^{t,\tau,i} = \lim_{N \rightarrow \infty} \inv{N} \nabla_{\bq_{\tau}} \cdot \Big( \bA^T \sum_{\tau' = 0}^{t} \sum_{k = 0}^{(t - \tau') i} r_{\tau'}^{t,i}[k] \bPhi^k \bz_{\tau'} \Big) \nonumber\\
    &= -\lim_{N \rightarrow \infty} \inv{N} Tr \bigg\{ \bA^T \sum_{k = 0}^{(t - \tau) i} r_{\tau}^{t,i}[k] \bPhi^k \bA \bigg\} \nonumber\\
    &= -\lim_{N \rightarrow \infty} \inv{N} Tr \bigg\{  \sum_{k = 0}^{(t - \tau) i} r_{\tau}^{t,i}[k] \bU \bLambda^{k+1} \bU^T \bigg\} \nonumber\\
    &= -\sum_{k = 0}^{(t - \tau) i} r_{\tau}^{t,i}[k] \lim_{N \rightarrow \infty} \inv{N} Tr \big\{ \bLambda^{k+1} \big\} = -\sum_{k = 0}^{(t - \tau) i} r_{\tau}^{t,i}[k] \chi_{k+1}
\end{align*}

\noindent
where we used the fact that $Tr\{ \bU \bLambda^k \bU^T \} = Tr\{\bLambda^k\}$ for any orthonormal matrix $\bU$ and the definition \eqref{eq:chi} of the spectral moments $\chi_k$.

Next, to work out the result \eqref{eq:v_ab_WS_CG} for the variance $\vAB^{t,i}$, we only need to show that $\inv{N} \big| \big|\bA^T \bmu_A^{t,i} \big| \big|^2 = \Omega_t^i$, where $\Omega_t^i$ is as in \eqref{eq:Omege}. Then, by using \eqref{eq:bz_tau_to_bmu}, we can obtain

\noindent
\begin{align}
    &\lim_{N \rightarrow \infty} \inv{N} \big| \big| \bA^T \bmu_A^{t,i} \big| \big|^2 \nonumber\\
    &= \lim_{N \rightarrow \infty} \inv{N} \bigg| \bigg| \bA^T \sum_{\tau = 0}^{t-1} \sum_{k = 0}^{(t - \tau) i} r_{\tau}^{t,i}[k] \bLambda^k (\bw - \bA \bq_t) \bigg| \bigg|^2 \nonumber\\
    &\as \lim_{N \rightarrow \infty} \inv{N} \bigg| \bigg| \bA^T \sum_{\tau = 0}^{t-1} \sum_{k = 0}^{(t - \tau) i} r_{\tau}^{t,i}[k] \bLambda^k \bA \bq_t \bigg| \bigg|^2 \nonumber\\
    &+ \inv{N} \bigg| \bigg| \bA^T \sum_{\tau = 0}^{t-1} \sum_{k = 0}^{(t - \tau) i} r_{\tau}^{t,i}[k] \bLambda^k \bw \bigg| \bigg|^2 \label{eq:A_mu_norm_step_a} \\
    &\as \limN \sum_{\tau,\tau' = 0}^{t-1} \sum_{j = 0}^{(t - \tau) i} \sum_{k = 0}^{(t - \tau') i'} r_{\tau}^{t,i}[k] r_{\tau'}^{t,i}[j] \frac{\bq_{\tau'}^T \bA^T \bPhi^{k+j+1} \bA \bq_{\tau}}{N} \nonumber\\
    &+ \sum_{\tau,\tau' = 0}^{t-1} \sum_{j = 0}^{(t - \tau) i} \sum_{k = 0}^{(t - \tau') i'} r_{\tau}^{t,i}[k] r_{\tau'}^{t,i}[j] \frac{\bw^T \bPhi^{k+j+1} \bw}{N} \label{eq:A_mu_norm_step_b}
\end{align}

\noindent
where in \eqref{eq:A_mu_norm_step_a} we used \eqref{eq:w_Aq_independence}. Next, by defining an auxiliary function $\bg(\bq_{\tau}) = \bPhi^{k+j+1} \bA \bq_{\tau}$ and using \eqref{eq:WS_gamma_A_equality} together with \eqref{eq:WS_gamma_A_assymptotic}, we can show that 

\noindent
\begin{align}
    \limN \inv{N} \bq_{\tau'}^T\ \bA^T \bg(\bq_{\tau}) &\as \psi_{\tau',\tau} \lim_{N \rightarrow \infty} \inv{N} \nabla_{\bq_{\tau}} \cdot \Big( \bA^T \bg(\bq_{\tau})  \Big) \nonumber\\
    &= \psi_{\tau',\tau} \lim_{N \rightarrow \infty} \inv{N} Tr \Big\{ \bA^T \bPhi^{k+j+1} \bA \Big\} \nonumber\\
    &= \psi_{\tau',\tau} \chi_{k+j+2} \label{eq:q_A_bg}
\end{align}

\noindent
Additionally, since $\bw$ is a zero-mean Gaussian vector, we can use the Stein's Lemma \cite{SURE} to get

\noindent
\begin{align}
    \limN \inv{N} \bw^T \bPhi^{k+j+1} \bw &\as v_w \limN \inv{N} Tr\big\{\bPhi^{k+j+1}\big\} \nonumber\\
    &= v_w \chi_{k+j+1} \label{eq:w_Phi_w}
\end{align}

\noindent
Substituting \eqref{eq:q_A_bg} and \eqref{eq:w_Phi_w} into \eqref{eq:A_mu_norm_step_b} gives the result as in \eqref{eq:v_ab_WS_CG}. \end{proof}

\subsection*{Asymptotics of $a_t^j$ and $b_t^j$ in WS-CG-VAMP}
 
Just above we confirmed the asymptotic result \eqref{eq:v_ab_WS_CG}, but for it to be a proper SE, we need all the components involved to be functions of either $v_w$, $\chi_j$ or $\psi_{\tau,\tau'}$. In \eqref{eq:v_ab_WS_CG} we have the function $\Omega_t^i$ that depends on $\br_{\tau}^{t,i}$, which is a function of $f_{\bmu}^{t,i}[j]$ (and other similar functions). This function is formulated in terms of $v_w$ and $\vBA^t$, but also in terms of $\{a_t^j\}_{j=0}^i$ and $\{b_t^j\}_{j=0}^i$ from the CG algorithm. Therefore, to complete the derivation of the SE, we need to define the asymptotic behaviour of these two sets of scalars. We begin with 

\noindent
\begin{equation}
    a_t^i = \frac{||\br_t^i ||^2}{(\bp_t^i)^T \bW_t \bp_t^i} = \frac{||\bz_t - \bW_t \bmu_A^{t,i} ||^2}{ (\bp_t^i)^T \bW_t \bp_t^i}\label{eq:a_t_i}
\end{equation}

\noindent
where we used the definition of the residual vector $\br_t^i$ from the CG algorithm. Then the norm in the numerator of \eqref{eq:a_t_i} corresponds to

\noindent
\begin{align}
    &\lim_{N \rightarrow \infty} \inv{N} ||\bz_t - \bW_t \bmu_A^{t,i} ||^2 \nonumber\\
    &= \lim_{N \rightarrow \infty} \inv{N} \big(||\bz_t||^2 - 2 \bz_t^T \bW_t \bmu_A^{t,i} + ||\bW_t \bmu_A^{t,i}||^2\big) \nonumber\\
    &= E_{1,t}^i - 2 E_{2,t}^i + E_{3,t}^i = E_t^i
\end{align}

\noindent
where we can use \eqref{eq:w_Aq_independence} and the fact that $\inv{N} Tr\{\bA \bA^T\} = 1$ to show that $E_{1,t}^i = \lim_{N \rightarrow \infty} \inv{N} ||\bz_t||^2 \overset{a.s.}{=} \delta v_w + \vBA^t$. Next, for $t = 0$ we can use \eqref{eq:CG_p_conjugate} and the fact that $\bp_0^0 = \bz_0$ to show that $\bz_t^T \bW_t \bmu_A^{t,i} = ||(\bz_0)||^2$ so

\noindent
\begin{equation}
    E_{2,0}^i = \limN \inv{N} ||(\bz_0)||^2 \as \delta v_w + \vBA^0
\end{equation}

\noindent
For $t>0$, this result does not hold due to the initializations \eqref{eq:mu_init} and \eqref{eq:p_init}. Instead, we use \eqref{eq:bz_tau_to_bmu} and \eqref{eq:w_Aq_independence} to obtain 

\noindent
\begin{align*}
    E_{2,t}^i &= \lim_{N \rightarrow \infty} \inv{N} \bz_t^T \bW_t \bmu_A^{t,i} \nonumber\\
    &= \lim_{N \rightarrow \infty} \inv{N} \sum_{\tau = 0}^{t-1} \sum_{k = 0}^{(t - \tau) i} r_{\tau}^{t,i}[k] \bz_t^T \bW_t \bPhi^k \bz_{\tau} \nonumber
\end{align*}

\noindent
Next we can use \eqref{eq:W_t}, \eqref{eq:z_t} and follow the same steps as in \eqref{eq:q_A_bg} and \eqref{eq:w_Phi_w} to show that

\noindent
\begin{align}
    &\limN \inv{N} \bz_t^T \bW_t \bPhi^k \bz_{\tau} \nonumber\\
    &\as \limN \inv{N} \Big( \bw^T \bW_t \bPhi^k \bw + \bq_t^T \bW_t \bPhi^k \bq_{\tau} \Big) \nonumber\\
    &\as v_w^2 \chi_k + v_w \psi_{t,\tau} \chi_{k+1} + (\psi_{t,\tau})^2 \chi_{k+2} = \Xi_{t,\tau}^k
\end{align}

\noindent
So that $E_{2,t}^i$ is equal to

\noindent
\begin{equation}
    E_{2,t}^i \as \limN \sum_{\tau = 0}^{t-1} \sum_{k = 0}^{(t - \tau) i} r_{\tau}^{t,i}[k] \Xi_{t,\tau}^k
\end{equation}

\noindent
for $t>0$. Following exactly the same steps as above, we can show that the norm $\limN \inv{N} ||\bW_t \bmu_A^{t,i}||^2$ almost surely converges to

\noindent
\begin{align*}
    E_{3,t}^i &= \lim_{N \rightarrow \infty} \inv{N} ||\bW_t \bmu_A^{t,i}||^2 \nonumber\\
    &\as \sum_{\tau, \tau' = 0}^{t-1} \sum_{j = 0}^{(t - \tau) i} \sum_{k = 0}^{(t - \tau') i} r_{\tau}^{t,i}[j] r_{\tau'}^{t,i}[k] \Delta_{\tau,\tau'}^{j+k}
\end{align*}

\noindent
where

\noindent
\begin{align*}
    \Delta_{\tau,\tau'}^j = v_w^3 \chi_j &+ 3 v_w^2 \psi_{\tau,\tau'} \chi_{j+1} + 3 v_w (\psi_{\tau,\tau'})^2 \chi_{j+2} \nonumber\\
    &+ (\psi_{\tau,\tau'})^3 \chi_{j+3}
\end{align*}

In the same way, we can expand the inner-product in the denominator of $a_t^j$ to obtain

\noindent
\begin{align*}
    L_t^i &= \lim_{N \rightarrow \infty} \inv{N} (\bp_t^i)^T \bW_t \bp_t^i \nonumber\\
    &\overset{a.s.}{=} \sum_{\tau, \tau' = 0}^{t-1} \sum_{j = 0}^{(t - \tau) i} \sum_{k = 0}^{(t - \tau') i} r_{\tau}^{t,i}[j] r_{\tau'}^{t,i}[k] \Upsilon_{\tau,\tau'}^{j+k}
\end{align*}

\noindent
with

\noindent
\begin{equation*}
    \Upsilon_{\tau,\tau'}^{j} = v_w^2 \chi_j + 2 v_w \psi_{\tau,\tau'} \chi_{j+1} + (\psi_{\tau,\tau'})^2 \chi_{j+2}
\end{equation*}

\noindent
As a result, we can show that $a_t^i$ and $b_t^j$ almost surely converge to

\noindent
\begin{equation}
    \lim_{N \rightarrow \infty} a_t^i \overset{a.s.}{=} \frac{E_t^i}{L_t^i} \quad \quad \quad \lim_{N \rightarrow \infty} b_t^i \overset{a.s.}{=} \frac{E_t^{i+1}}{E_t^i}
\end{equation}

\noindent
which are function of $v_w$, $\chi_j$ and $\psi_{\tau,\tau'}$ only.

\section*{Appendix D}

\begin{proof}[\unskip\nopunct]

Next, we prove Theorem \ref{theorem:alternative_gamma_t_tau_estimator}. Recall from Section \ref{section:WS_CG_iterative_gamma} that $(\bgamma_A^{t,i})_{\tau} = \gamma_A^{t,\tau,i}$ can be recovered from

\noindent
\begin{equation*}
    \bgamma_A^{t,i} \as \lim_{N \rightarrow \infty} \bPsi_t^{-1} \big( \nu_t^i \bone_t - \inv{N} \bZ_t^T \bmu_A^{t,i}\big)
\end{equation*}

\noindent
provided the asymptotic result of the inner-product $\nu_t^i = \inv{N} \bw^T \bmu_A^{t,i}$. Note that based on the definition of $\bmu_A^{t,i}$ from Algorithm 2, we have for $i>0$ that 

\noindent
\begin{align}
    \limN \nu_t^i &= \limN \inv{N} \bw^T \Big( \bmu_A^{t,i-1} + a_t^{i-1} \bp_t^{i-1} \Big) \nonumber\\
    &= \limN \nu^{i-1} + a_t^{i-1} \eta_t^{i-1} \label{eq:appendix_c_barpsi_derived}
\end{align}

\noindent
where we defined $\eta_t^i = \inv{N} \bw^T \bp_t^i$, which can be further expanded using the definition of $\bp_t^i$ from Algorithm 2 as

\noindent
\begin{align}
    \limN \eta_t^i &= \lim_{N \rightarrow \infty} \inv{N} \bw^T \bp_t^i \nonumber\\
    &= \lim_{N \rightarrow \infty} \inv{N} \bw^T \Big( \bz_t - \bW_t \bmu_A^{t,i} + b_t^{i-1} \bp_t^{i-1} \Big) \nonumber\\
    &\overset{a.s.}{=} \delta v_w + \limN b_t^{i-1} \eta_t^{i-1} - \inv{N} \bw^T \bW_t \bmu_A^{t,i}\label{eq:appendix_c_bareta_expansion}
\end{align}

\noindent
where we used \eqref{eq:w_Aq_independence} to show that $\lim_{N \rightarrow \infty} \inv{N} \bw^T \bz_t \overset{a.s.}{=} \delta v_w$. Additionally, by expanding the matrix $\bW_t$, we arrive at

\noindent
\begin{equation}
    \limN \frac{\bw^T \bW_t \bmu_A^{t,i}}{N} = v_w \nu_t^i + \vBA^t \limN \frac{\bw^T \bPhi^T \bmu_A^{t,i}}{N} \label{eq:w_W_mu}
\end{equation}

\noindent
Next, by using \eqref{eq:bz_tau_to_bmu} we can show that

\noindent
\begin{align}
    &\limN \inv{N} \bw^T \bPhi \bmu_A^{t,i} = \limN \sum_{\tau = 0}^{t} \sum_{k = 0}^{(t - \tau) i} r_{\tau}^{t,i}[k] \inv{N} \bw^T \bPhi \bPhi^k \bz_{\tau} \nonumber\\
    &\as \sum_{\tau = 0}^{t} \sum_{k = 0}^{(t - \tau) i} r_{\tau}^{t,i}[k] \limN \inv{N} \bw^T \bPhi^{k+1} \bw \label{eq:w_bPhi_mu_1} \\
    &\as v_w \sum_{\tau = 0}^{t} \sum_{k = 0}^{(t - \tau) i} r_{\tau}^{t,i}[k] \limN \inv{N} Tr\big\{ \bPhi^{k+1}\big\} \label{eq:w_bPhi_mu_2}\\
    &= v_w \sum_{\tau = 0}^{t} \sum_{k = 0}^{(t - \tau) i} r_{\tau}^{t,i}[k] \chi_{k+1} \overset{(a)}{=} - v_w \sum_{\tau = 0}^{t} \gamma_A^{t,\tau,i} = - v_w ||\bgamma_A^{t,i}||_1  \label{eq:w_bPhi_mu_4}
\end{align}

\noindent
where \eqref{eq:w_bPhi_mu_1} follows from \eqref{eq:w_Aq_independence}, \eqref{eq:w_bPhi_mu_2} is due to the Stain's Lemma \cite{SURE} and \eqref{eq:WS_CG_correction_scalars} motivates the step (a). Then, by combining  \eqref{eq:appendix_c_barpsi_derived}, \eqref{eq:appendix_c_bareta_expansion}, \eqref{eq:w_bPhi_mu_4} and \eqref{eq:Gamma_t_i_WS_CG}, we arrive at the following recursion

\noindent
\begin{gather*}
    \nu_t^i = \nu^{i-1} + a_t^{i-1} \eta_t^{i-1} \\
    \bgamma_A^{t,i} \as \lim_{N \rightarrow \infty} \bPsi_t^{-1} \big( \nu_t^i \bone_t - \inv{N} \bZ_t^T \bmu_A^{t,i}\big) \\
    \eta_t^i = v_w \big(\delta - \nu_t^i + \vBA^t ||\bgamma_A^{t,i}||_1\big) + b_t^{i-1} \eta_t^{i-1}
\end{gather*}

Lastly, we consider the initialization of the recursion. For $t>0$ we have $\bmu_A^{t,0} = \bmu_A^{t-1,i}$ so that

\noindent
\begin{equation}
    \nu_t^0 =  \inv{N} \bw^T \bmu_A^{t,0} = \inv{N} \bw^T \bmu_A^{t-1,i} = \nu_{t-1}^i
\end{equation}

\noindent
Next, recall that we initialize $\bp_t^0 = \bz_t - \bW_t \bmu_A^{t-1,i} + b_{t-1}^{i-1} \bp_{t-1}^{i-1}$. Then we have

\noindent
\begin{align*}
    \limN \eta_t^0 &= \limN \inv{N} \bw^T \bp_t^0 \nonumber\\
    &= \limN \inv{N} \bw^T \big( \bz_t - \bW_t \bmu_A^{t-1,i} + b_{t-1}^{i-1} \bp_{t-1}^{i-1} \big) \nonumber\\
    &\as \delta v_w - v_w \nu_{t-1}^i + \vBA^t v_w ||\bgamma_A^{t-1,i}||_1 + b_{t-1}^{i-1} \eta_{t-1}^{i-1} \nonumber\\
    &= v_w \big(\delta - \nu_{t-1}^i + \vBA^t ||\bgamma_A^{t-1,i}||_1 \big) + b_{t-1}^{i-1} \eta_{t-1}^{i-1}
\end{align*}

\noindent
where we used \eqref{eq:w_W_mu} and \eqref{eq:w_bPhi_mu_4} to show that

\noindent
\begin{align*}
    \limN \inv{N} \bw^T \bW_t \bmu_A^{t-1,i} &= v_w \nu_{t-1} + \vBA^{t} \limN \frac{\bw^T \bPhi^T \bmu_A^{t-1,i}}{N} \nonumber\\
    &\as v_w \nu_{t-1} - \vBA^{t} v_w ||\bgamma_A^{t-1,i}||_1
\end{align*}

For $t=0$, we have $\bmu_{-1}^i = \bp_{-1}^i = \bzero_M$ so $\nu_0^0 =  \inv{N} \bw^T \bmu_0^0 = 0$. Lastly, in the limit, we initialize $\limN \eta_0^0 = \limN \inv{N} \bw^T \bp_0^0 = \limN \inv{N} \bw^T \bz_0 \overset{a.s.}{=} \delta v_w$, which completes the proof. \end{proof}

\subsection*{Proving Theorem \ref{theorem:zero_init_gamma_estimator}} 

\begin{proof}[\unskip\nopunct]

In this subsection we utilize the proof approach for Theorem \eqref{theorem:alternative_gamma_t_tau_estimator} to prove Theorem \ref{theorem:zero_init_gamma_estimator}. Recall from Section \ref{sec:stable_estimation_in_CG_VAMP} that for the zero-initialized CG we need a single correction scalar $\gamma_A^{t,i}$ that follows 

\noindent
\begin{equation}
    \gamma_A^{t,i} \as \limN \frac{\inv{N} \bw^T \bmu_A^{t,i} - \inv{N} \bz_t^T \bmu_A^{t,i}}{\vBA^t} \label{eq:gamma_A_zero_init_CG_proof}
\end{equation}

\noindent
As for WS-CG case, we need to find an asymptotic identity for $\limN \nu_t^i = \limN \inv{N} \bw^T \bmu_A^{t,i}$ that we could estimate iteratively. Since the iterations of WS-CG and of zero-initialized CG are the same, the steps \eqref{eq:appendix_c_barpsi_derived} - \eqref{eq:w_W_mu} are the same for both algorithm as well. What changes is the asymptotic result of the inner-product $\inv{N} \bw^T \bPhi \bmu_A^{t,i}$ that now is equal to

\noindent
\begin{align}
    \limN &\inv{N} \bw^T \bPhi \bmu_A^{t,i} \overset{(a)}{=} \sum_{j=0}^i r_t^{t,i}[j] \limN \inv{N} \bw^T \bPhi^{j+1}  \bz_t \nonumber\\
    &\as \sum_{j=0}^i r_t^{t,i}[j] \limN \inv{N} \bw^T \bPhi^{j+1}  \bw \label{eq:w_bPhi_zero_init_mu_3}\\
    &\as v_w  \sum_{j=0}^i r_t^{t,i}[j] \limN \inv{N} Tr \big\{ \bPhi^{j+1} \big\} \label{eq:w_bPhi_zero_init_mu_1}\\
    &\overset{(b)}{=} v_w \sum_{j=0}^i r_t^{t,i}[j] \chi_{j+1} \overset{(c)}{=} - v_w \gamma_A^{t,i}  \label{eq:w_bPhi_zero_init_mu_2}
\end{align}

\noindent
where in (a) we used \eqref{eq:zero_init_CG_LSL_model}, \eqref{eq:w_bPhi_zero_init_mu_3} follows from \eqref{eq:w_Aq_independence}, \eqref{eq:w_bPhi_zero_init_mu_1} is due to to the Stein's Lemma \cite{SURE}, in (b) we used \eqref{eq:chi} and (c) follows from \eqref{eq:zero_init_CG_gamma_A}. Next, substituting \eqref{eq:w_bPhi_zero_init_mu_2} into \eqref{eq:w_W_mu} gives

\noindent
\begin{align}
    \limN &\frac{\bw^T \bW_t \bmu_A^{t,i}}{N} = \limN v_w  \nu_t^i - \vBA^t v_w \gamma_A^{t,i} \nonumber\\
    &\as \limN v_w \nu_t^i + v_w \Big( \inv{N} \bz_t^T \bmu_A^{t,i} - \inv{N} \bw^T \bmu_A^{t,i} \Big) \label{eq:w_W_zero_init_mu_1}\\
    &= v_w \limN \inv{N} \bz_t^T \bmu_A^{t,i} \label{eq:w_W_mu_zer_init_CG_proof}
\end{align}

\noindent
where \eqref{eq:w_W_zero_init_mu_1} comes from \eqref{eq:gamma_A_zero_init_CG_proof}. Then \eqref{eq:w_W_mu_zer_init_CG_proof} together with \eqref{eq:appendix_c_bareta_expansion} implies

\noindent
\begin{align}
    \limN \eta_t^i &\overset{a.s.}{=} \delta v_w + \limN b_t^{i-1} \eta_t^{i-1} - v_w \inv{N} \bz_t^T \bmu_A^{t,i} \nonumber\\
    &= \limN v_w \big(\delta - \inv{N} \bz_t^T \bmu_A^{t,i} \big) + b_t^{i-1} \eta_t^{i-1} \label{eq:eta_iterative}
\end{align}

\noindent
which matches the iterations \eqref{eq:LSL_gamma_A_rec_1}. Additionally, by reapplying \eqref{eq:appendix_c_barpsi_derived} to itself $i$ times, we arrive at

\noindent
\begin{equation}
    \limN \inv{N} \bw^T \bmu_A^{t,i} \as \limN \sum_{j=0}^i a_t^j \eta_t^j
\end{equation}

\noindent
Lastly, based on the definition of $\eta_t^i$, we have the following limiting initialization $\limN \eta_t^0 = \limN \inv{N} \bw^T \bp_t^0 = \limN \inv{N} \bw^T \bz_t \overset{a.s.}{=} \delta v_w$, which completes the proof. \end{proof}

\section*{Appendix E}
\begin{proof}[\unskip\nopunct]

Next, we prove Theorem \ref{theorem:zero_init_CG_v_ab_efficient} that provides an alternative version of the limit

\noindent
\begin{equation}
    \vAB^{t,i} \as (\gamma_A^{t,i})^{-2} \limN \inv{N} ||\bA^T \bmu_A^{t,i}||^2 - \vBA^t \label{eq:v_AB_practical_but_slow_proof}
\end{equation}

\noindent
For this, we define $\bPhi = \bA \bA^T$ and note that \eqref{eq:W_t} implies

\noindent
\begin{equation}
    \bPhi =  \frac{ \bW_t - v_w \bI_M }{\vBA^t} \label{eq:bPhi}
\end{equation}

\noindent
Then, by expanding the norm in \eqref{eq:v_AB_practical_but_slow_proof} and using \eqref{eq:bPhi} gives

\noindent
\begin{equation}
    \vAB^{t,i} \as \limN \frac{ \big( \bmu_A^{i,t} \big)^T \bW_t \bmu_A^{i,t} - v_w ||\bmu_A^{i,t}||^2 }{N \vBA^t (\gamma_A^{t,i})^2} - \vBA^t \label{eq:v_AB_practical_intermediate_proof}
\end{equation}

\noindent
As discussed in Section \ref{sec:CG_introduction}, the CG output $\bmu_A^{t,i}$ can be represented as a linear combination of the conjugate direction vectors

\noindent
\begin{equation*}
    \bmu_A^{i,t} = \sum_{j = 0}^i a_t^j \bp_t^j
\end{equation*}

\noindent
where $\big( \bp_t^i \big)^T \bW_t \bp_t^j = 0 $ for any $i \neq j$. Thus, the inner product $\big( \bmu_A^{i,t} \big)^T \bW_t \bmu_A^{i,t}$ can be equivalently represented as

\noindent
\begin{equation}
    \big( \bmu_A^{i,t} \big)^T \bW_t \bmu_A^{i,t} = \sum_{j = 0}^i \big( a_t^j \big)^2 \big( \bp_t^j \big)^T \bW_t \bp_t^j \label{eq:mu_W_mu}
\end{equation}

\noindent
Additionally, using the definition of $a_t^j = \frac{||\br_t^j||^2}{ (\bp^j)^T \bW_t \bp_t^j}$ from Algorithm 2, we can further simplify \eqref{eq:mu_W_mu} to

\noindent
\begin{equation}
    \big( \bmu_A^{i,t} \big)^T \bW_t \bmu_A^{i,t} = \sum_{j = 0}^i  a_t^j  ||\br_t^j||^2
\end{equation}

\noindent
Then, by defining $\zeta_t^i = \zeta_t^{i-1} + a_t^i ||\br_t^i||^2$ with $\zeta_t^0 = 0$ and substituting these results back into \eqref{eq:v_AB_practical_intermediate_proof} completes the proof. \end{proof}

\printbibliography

\begin{IEEEbiographynophoto}{Nikolajs Skuratovs} 
received the B.Eng degree in electrical engineering, in 2017, from the Transport and Telecommunication Institute (TSI), Riga, Latvia, and the M.Sc. degree in signal processing and communications, in 2018, from The University of Edinburgh (UoE), Edinburgh, U.K., where he is currently working towards the Ph.D. degree. 
\end{IEEEbiographynophoto}

\begin{IEEEbiographynophoto}{Mike E. Davies} 
(Fellow, IEEE) received the M.A. degree in engineering from Cambridge University, Cambridge, U.K., in 1989, and the Ph.D. degree in nonlinear dynamics from University College London (UCL), London, U.K., in 1993. 

He was the Head of the Institute for Digital Communications (IDCOM), The University of Edinburgh (UoE), Edinburgh, U.K., from 2013 to 2016. He holds the Jeffrey Collins Chair in signal and image processing at UoE, where he also leads the Edinburgh Compressed Sensing Research Group. He was awarded a Royal Society University Research Fellowship in 1993 and was a Texas Instruments Distinguished Visiting Professor at Rice University, Houston, TX, USA, in 2012. He leads the U.K. University Defence Research Collaboration (UDRC) Program on signal processing for defense in collaboration with the U.K. Defence Science and Technology Laboratory (DSTL). His research has focused on nonlinear time series, source separation, compressed sensing and computational imaging. He has also explored the application of these ideas to advanced medical imaging, RF sensing applications, and machine learning. 

Prof. Davies has been elected as a fellow of EURASIP, the Royal Society of Edinburgh, and the Royal Academy of Engineering. He was a recipient of the European Research Council (ERC) Advanced Grant on Computational Sensing and the Royal Society Wolfson Research Merit award. He was awarded the Foundation Scholarship for the M.A. degree in 1987.
\end{IEEEbiographynophoto}

\end{document}